\documentclass[useAMS,usenatbib,usegraphicx]{mn2e}

\title[Influence of the Galactic bar]
{Influence of the Galactic bar on the kinematics of the disc stars
with {\it Gaia} EDR3 data} \author[Melnik et
al.]{A.M.~Melnik$^1$\thanks{E-mail: anna@sai.msu.ru}, A.K.~
Dambis$^1$, E.N.~Podzolkova$^2$, L.N.~Berdnikov$^1$,\\
$^1$Sternberg Astronomical Institute, Lomonosov Moscow State
University, Universitetskii pr. 13, Moscow, 119991, Russia \\
$^2$Faculty of Physics, Lomonosov Moscow State University, Leninskie
Gory 1-2, Moscow, 119991, Russia  }

\begin{document}

\date{Accepted 2021 December 00. Received 2021 December 00; in original form 2021 December 00}

%\pagerange{\pageref{firstpage}--\pageref{lastpage}} \pubyear{2009}

\maketitle

\label{firstpage}

\begin {abstract}

A model of the Galaxy with the outer ring $R_1R_2$ can explain the
observed distribution of the radial, $V_R$, and azimuthal, $V_T$,
velocity components along the Galactocentric distance, $R$, derived
from the {\it Gaia} EDR3 data. We selected stars from the {\it Gaia}
EDR3 catalogue with reliable parallaxes, proper motions and
line-of-sight velocities lying near the Galactic plane, $|z|<200$ pc,
and in the sector of the Galactocentic angles $|\theta|<15^\circ$ and
calculated the median velocities $V_R$ and $V_T$ in  small bins along
the distance $R$. The distribution of observed velocities appears to
have some specific features:  the radial velocity $V_R$ demonstrates
a smooth fall from   +5 km s$^{-1}$ at the distance of $R \approx
R_0-1.5$ kpc to  $-3$ km s$^{-1}$ at $R \approx R_0+1.0$ kpc while
the azimuthal velocity $V_T$ shows a sharp drop by  7 km s$^{-1}$ in
the distance interval  $R_0<R<R_0+1.0$ kpc, where $R_0$ is the solar
Galactocentric distance. We build a model of the Galaxy including
bulge, bar, disc and halo components, which reproduces the observed
specific features of the velocity distribution in the Galactocentric
distance interval $|R-R_0|< 1.5$ kpc. The best agreement corresponds
to the time $1.8\pm0.5$ Gyr after the start of the simulation. A
model of the Galaxy with the bar rotating at the angular velocity of
$\Omega_b=55\pm3$ km s$^{-1}$ kpc$^{-1}$, which sets the OLR of the
bar at the distance of $R_0-0.5\pm0.4$ kpc, provides the best
agreement between the model and observed velocities. The  position
angle of the bar, $\theta_b$, corresponding to the best agrement
between the model and observed velocities is
$\theta_b=45\pm15^\circ$.

\end{abstract}


\section{Introduction}

Bars were found in nearly  70 per cent of bright disc galaxies
\citep{eskridge2000, menendez2007}. On the whole, the fraction of
barred galaxies increases with  decreasing redshift, but during the
last $\sim 4$ Gyr, the fraction of barred galaxies changes negligibly
\citep{sheth2008,melvin2013}. In many cases the formation of a bar
does not require an external perturbation but results from the
secular evolution of galaxies \citep{kormendy2004}.

There are a lot of data indicating the presence of the bar in the
Galaxy. The gas kinematics  \citep{pohl2008, gerhard2011,
pettitt2014}, infrared observations \citep{dwek1995, benjamin2005,
cabrera-lavers2007, churchwell2009, gonzalez2012}, as well as the
X-shaped distribution of red giants in the central part of the disc
\citep{li2012, nesslang2016, simion2017} confirm the presence of a
bar in the Galaxy. Estimates of the length of the bar semi-major axis
and its angular velocity lie in the range $a=3$--5 kpc and
$\Omega_b=30$--60 km s$^{-1}$ kpc$^{-1}$, respectively. The position
angle of the bar with respect to the Sun is supposed to be
$\theta_b=15\textrm{--}45^\circ$, implying that the end of the bar
that is nearest to the Sun is located in quadrant I.

The age of the Galactic bar is a subject of scientific debates
\citep{nataf2016, haywood2016, bensby2017, bernard2018, fujii2019,
carrillo2019}. Here we consider a strong bar rotating with a constant
angular velocity, though  oval structures existed in the Galactic
disc much earlier. The presence of young metal-rich stars in the
Galactic  disc at the distances of $R=2$--3 kpc from the center
supports the idea that the Galactic bar  formed 2--3 Gyr ago
\citep[][and other papers]{debattista2019, baba2020,
hasselquist2020}.

The locations of the Outer and Inner Lindblad Resonances of the bar
are determined by the ratio of the difference between the angular
velocity of the bar and the disc $(\Omega-\Omega_b)$ to the frequency
of the epicyclic motion $\kappa$:

\begin{equation}
\frac{\kappa}{\Omega-\Omega_b}=\pm2/1,
\end{equation}

\noindent where $-2/1$ corresponds to the Outer  Lindblad Resonance
(OLR), but +2/1 -- to the Inner Lindblad Resonance (ILR). Besides,
resonances of order $\pm 4/1$ are also important
\citep{athanassoula1992,contopoulos1980, contopoulos1989}.

The resonance between the orbital rotation with respect to the bar
and the epicyclic motion causes the formation of  elliptic resonance
rings. There are three types of resonance rings: nuclear ($n$), inner
($r$) and outer ($R_1$ and $R_2$) rings. The nuclear rings lie near
the ILR  and are stretched perpendicular to the bar; the inner rings
form near the inner 4/1 resonance and are oriented along the bar; the
outer rings $R_1$ and $R_2$ are located near the OLR: the rings $R_1$
or pseudorings $R_1'$ (broken rings) lie a bit closer to the galactic
center and are stretched perpendicular to the bar while the rings
$R_2$ or pseudorings $R_2'$ are located a bit father away from the
center and are stretched parallel to the bar. There is also a mixed
morphological type $R_1R_2'$ including both $R_1$ and $R_2'$ outer
rings/pseudorings \citep{buta1995, buta1996, buta1991, rodriguez2008,
sormani2018}. The fraction of galaxies with outer rings is as high as
20--30 per cent among galaxies with strong and moderate bars
\citep{comeron2014}.

The backbone of resonance rings is stable direct periodic orbits
which are followed by numerous quasi-periodic orbits. There are two
basic families of stable direct periodic orbits, $x_1$ and $x_2$.
Orbits of the $x_1$ family support the bar inside the corotation
radius. Orbits of the $x_2$ family are elongated perpendicular to the
bar and support the nuclear rings between two ILRs. Near the OLR of
the bar the main family of periodic orbits $x_1$ splits into two
families: $x_1(1)$ and $x_1(2)$. The main stable periodic orbits
$x_1(2)$ lying between the  $-4/1$ and $-2/1$ (OLR) resonances are
elongated perpendicular to the bar while  the orbits $x_1(1)$ located
outside the OLR are stretched along the bar. The periodic orbits
$x_1(2)$ support the outer rings $R_1$ while the orbits $x_1(1)$
support the outer rings $R_2$ \citep{contopoulos1980,
contopoulos1989, schwarz1981, buta1996}.

\citet{athanassoula2009} studied manifold tubes  emanating from the
unstable Lagrangian points located near the ends of the bar. They
showed that less strong bars give rise to $R_1$ rings/pseudorings,
while stronger bars drive spirals, $R_2$ and $R_1R_2$
rings/pseudorings.

The bimodal velocity distribution of disc stars in the solar
neighborhood can be explained by the location of the Sun near the OLR
of the bar \citep[][and other papers]{kalnajs1991, dehnen2000,
fux2001, fragkoudi2019, sanders2019, asano2020, trick2021,
chiba2021}. Furthermore, galaxies with a flat rotation curve and a
fast bar, which ends near the corotation radius
\citep{debattista2000, rautiainen2008}, produce an OLR  at a distance
nearly twice exceeding the length of the bar semi-major axis,
$R\approx 2a$. For the Galaxy, this distance nearly corresponds to
the solar Galactocentric distance, $R_0$.

Modelling of the outer rings shows that they  form  0.5--1.0 Gyr
after the bar is turned on. The ring $R_1$ appears first while the
ring $R_2$ forms a bit later, $\sim 1$ Gyr after the start of the
simulation \citep{schwarz1981, byrd1994, rautiainen1999,
rautiainen2000}. \citet{rautiainen2000} studied the outer ring
morphology in N-body models and found cyclic changes in the ring
type: from $R_1R_2'$ to $R_2'$ and back to $R_1R_2'$.

Models with analytical bars are of particular interest in studies of
the kinematics of  Galactic stars. \citet{melnikrautiainen2009}
showed that a model of the Galaxy with a two-component outer ring
$R_1R_2'$ can explain the average velocities of young stars in the
Sagittarius and Perseus star-gas complexes \citep{efremov1988}.
\citet{melnik2019} found that a  model of the Galaxy with  a bar
rotating with the angular velocity of $\Omega_b=50\pm2$ km s$^{-1}$
kpc$^{-1}$  can reproduce the average velocities of young stars in
three star-gas complexes: Sagittarius, Perseus and Local System. The
Sagittarius complex with the average radial velocity,
$V_R=+7.5\pm2.1$ km s$^{-1}$,  directed away from the Galactic center
belongs to the ring $R_1$ while the Perseus complex with the radial
velocity, $V_R=-4.7\pm2.2$ km s$^{-1}$, directed toward the Galactic
center is associated with the ring $R_2$. The Local System is located
between these two complexes and has the average radial velocity of
$V_R=+5.4\pm2.6$ km s$^{-1}$. The model considered can reproduce the
velocity in the Local System during the time period 1--2 Gyr after
the start of modelling. The best agreement with observations
corresponds to the position angle of the bar lying in the range
$\theta_b=40\textrm{--}52^\circ$.

\citet{rautiainen2010} built an N-body model of the Galaxy which
demonstrates the formation of a bar and  outer rings. The velocities
of model particles were averaged  over 1 Gyr   time periods for
comparison with observations. Such averaging  suppresses the
influence of slow modes and random velocity changes. The average
model velocities in N-body models reproduce the observed velocities
of young stars in the Sagittarius, Perseus and Local System star-gas
complexes.

The kinematics and distribution of classical Cepheids, OB
associations and young star clusters in the 3-kpc solar neighborhood
are indicative of the presence of  "the tuning-fork-like" structure
in the Galactic disc, which can be accounted for by the existence of
two segments of the outer rings fusing together near the Carina
star-gas complex \citep{melnik2015, melnik2016}. In addition, models
of the Galaxy with a two-component outer ring $R_1R_2'$ can reproduce
the location of the Carina-Sagittarius spiral arm, where the Carina
arm is located near the segment of the outer ring $R_2$ while the
Sagittarius arm lies near the ring $R_1$ \citep{melnik2011}.

The early installment of the third  {\it Gaia} data release ({\it
Gaia} EDR3) including proper motions and parallaxes for 1.5 billion
stars  opens new possibilities for the study of the Galactic
structure and kinematics. {\it Gaia} EDR3 also lists 7.2 million
stellar line-of-sight velocities presented by the second {\it Gaia}
data release ({\it Gaia} DR2) and adopted by {\it Gaia} EDR3 with
some small corrections \citep{prusti2016, brown2018, katz2018,
brown2020, lindegren2020}.

In this paper we study the kinematics of  Galactic stars with the
{\it Gaia} EDR3  data and build a model of the Galaxy which
reproduces the observed distributions of the radial and azimuthal
velocities along the Galactocentric distance. Section 2 describes the
observational data. Section 3 presents the dynamical model of the
Galaxy. Section 4 describes the results: a comparison of the observed
and model velocities and  velocity dispersions, the search for the
optimal values of the bar angular velocity, $\Omega_b$, and the
positional angle of the bar,  $\theta_b$. The discussion and main
conclusions are given in section 5.

\section{Observational data}\label{nd}

The average accuracy of proper motions and parallaxes of the {\it
Gaia} EDR3 catalogue allows the velocities and distances to stars
located within  1 kpc from the Sun to be derived with the average
error of $\pm0.2$ km s$^{-1}$ and $\pm50$ pc, respectively. The
average accuracy of {\it Gaia} EDR3 line-of-sight velocities  is $\pm
2$ km s$^{-1}$ \citep{brown2020,brown2018}.

We selected  {\it Gaia} EDR3 stars located near the Galactic plane,
$|z|< 200$ pc, and   in the sector of the Galactocentric angles
$|\theta|<15^\circ$ that have parallaxes, $\varpi$, determined  with
the relative error less  than 20 per cent
($\varpi/\varepsilon_\varpi>5$) and line-of-sight velocities measured
by  the {\it Gaia} spectrometer. We excluded from the initial sample
of {\it Gaia} EDR3 stars  (2886715 stars) 493472 objects ($17$ per
cent) with the re-normalized error (RUWE) greater than
$\textrm{RUWE}>1.4$ \citep{lindegren2018}. The final sample of {\it
Gaia} EDR3 stars  includes 2393243 objects. The heliocentric
distances  to stars were derived from {\it Gaia} EDR3 parallaxes
without any zero-point correction: $r=1/\varpi$.

We also  compared observational distributions derived from the {\it
Gaia} EDR3 and {\it Gaia} DR2 data. The initial sample of {\it Gaia}
DR2 stars lying near the Galactic plane, $|z|< 200$ pc, and   in the
sector of the Galactocentric angles $|\theta|<15^\circ$ that have
reliable parallaxes and known line-of-sight velocities  includes
2987601 objects. We excluded from  the initial sample  243059 stars
($8.1$ per cent) with the re-normalized error $\textrm{RUWE}>1.4$ and
75504 stars ($2.5$ per cent) with the number of visibility periods
$n_{vis} \le 8$. The heliocentric distances, $r$, were derived from
{\it Gaia} DR2 parallaxes, $\varpi$,  in the following way:

\begin{equation}
r=\frac{1}{\varpi-\Delta \varpi},
\end{equation}

\noindent where $\Delta \varpi$ is a parallax  zero-point offset. The
value of the offset $\Delta \varpi$ depends on the  magnitude $G$:
the brighter the star, the larger the absolute value of $\Delta
\varpi$. For stars with $G=8^m$  the offset $\Delta \varpi$ attains
the value of $-0.120$ mas while for quasars it is only $\Delta
\varpi=-0.030$ mas \citep{arenou2018, lindegren2018, melnik2020}. We
excluded   stars with the magnitude $G<10^m$ (220710 stars) and
adopted for all {\it Gaia} DR2 stars  the parallax zero-point offset
of $\Delta \varpi=-0.05$ mas \citep{riess2018, zinn2019, leung2019,
yalyalieva2018, schonrich2019}. The final sample of {\it Gaia} DR2
stars  includes 2448328 objects.

\begin{figure*}
\resizebox{\hsize}{!}{\includegraphics{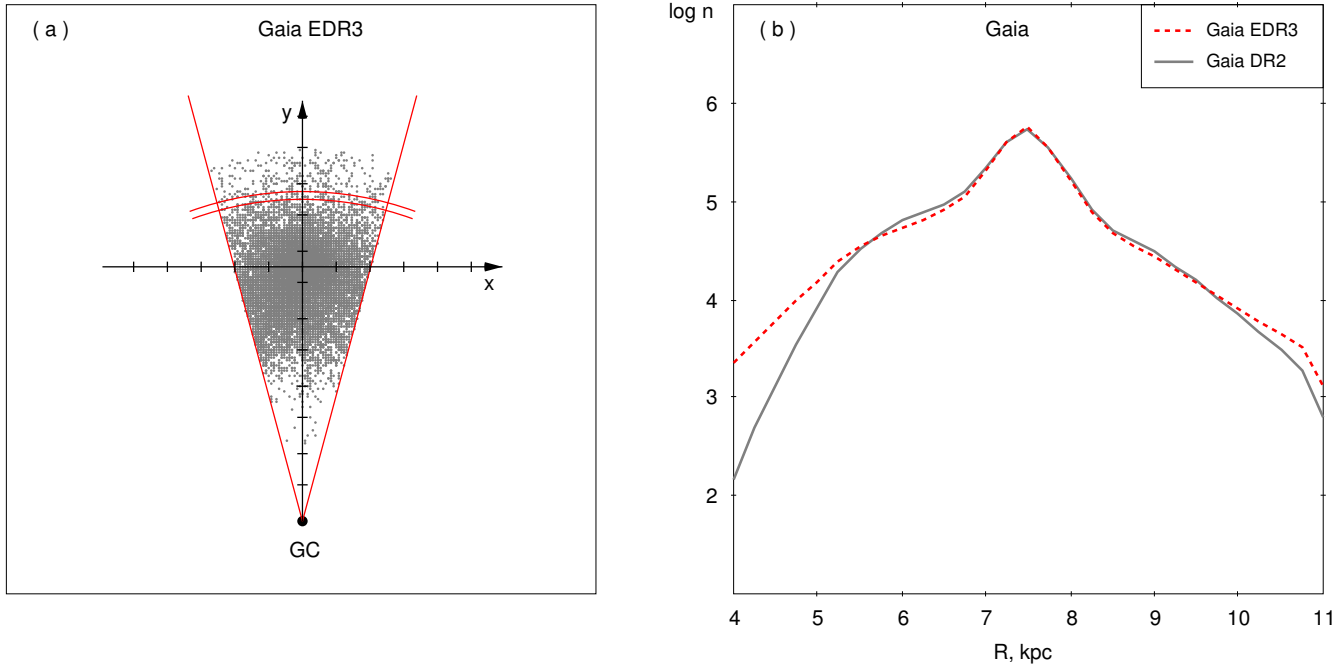}} \caption{ (a)
Distribution of  {\it Gaia} EDR3 stars   located near the Galactic
plane, $|z|< 200$ pc, and in the sector of the Galactocentric angles
$|\theta|<15^\circ$ (gray points). We selected {\it Gaia} EDR3 stars
with reliable parallaxes, known proper motions and line-of-sight
velocities. Only  1 per cent of objects are shown. For an example, we
outlined one  $\Delta R=250$-pc wide bin by two arcs. The Galactic
center (GC) is at the bottom, the axes $x$ and $y$ are directed
toward the Galactic rotation and away from the Galactic center,
respectively; the Sun is at the origin. (b) Logarithmic dependencies
of the numbers of stars, $n$, in  250-pc wide bins on the distance
$R$ derived from the {\it Gaia} EDR3 and {\it Gaia} DR2 data.}
\label{gaia_distrib}
\end{figure*}

Fig.~\ref{gaia_distrib}(a) shows the observational sample of stars
selected from the  {\it Gaia} EDR3 catalogue. The distribution of
stars along the Galactocentric distance $R$ was subdivided into
250-pc wide bins. Fig.~\ref{gaia_distrib}(b) shows the number of
stars, $n$, in bins at different distances $R$ calculated for the
{\it Gaia} EDR3 and {\it Gaia} DR2 catalogues. We can see that the
{\it Gaia} EDR3 catalogue includes considerably more stars in the
distance interval $R=4$--5 kpc than the {\it Gaia} DR2 catalogue.

We adopted the solar Galactocentric distance to be of $R_0=7.5$ kpc
\citep{glushkova1998, nikiforov2004, feast2008, groenewegen2008,
reid2009b, dambis2013, francis2014, boehle2016, branham2017}. On the
whole, the choice of the value of $R_0$ in the range 7--9 has
virtually no effect on the  results.

The velocity components of the Sun in the Galactic centre rest frame
in the directions  toward the Galactic rotation, $V_{T0}$, toward the
Galactic center, $V_{R0}$, and in the direction perpendicular to the
Galactic plane, $V_{Z0}$, are adopted to be $V_{T0}=\Omega_0
R_0+12.0$ km s$^{-1}$, $V_{R0}=10.0$ km s$^{-1}$ and $V_{Z0}=7.0$ km
s$^{-1}$, where $\Omega_0$ is the angular velocity of the rotation of
the Galactic disc at the solar distance $R_0$ and its value is taken
to be $\Omega_0=30.0$ km s$^{-1}$ kpc$^{-1}$. The adopted $\Omega_0$,
$V_{T0}$, $V_{R0}$  and $V_{Z0}$ are consistent with values derived
from an analysis of the kinematics of OB associations with the {\it
Gaia} DR2 data \citep{melnik2020}.

The velocities of stars in the  radial, $V_R$, and  azumuthal, $V_T$,
directions as well as in the direction perpendicular to the Galactic
plane, $V_Z$, are computed in the following way:

\begin{equation}
\begin{array}{c}
 V_R= (V_r\, \cos b - V_b\, \sin b)\; \sin \alpha + V_l\, \cos \alpha \\
 +V_{T0}\,\sin \theta  - V_{R0}\,\cos \theta,
\end{array}
\end{equation}

\begin{equation}
\begin{array}{c}
 V_T= (V_r\, \cos b - V_b\, \sin b)\; \cos \alpha - V_l\, \sin \alpha \\
 +V_{T0}\,\cos \theta + V_{R0}\,\sin \theta,
\end{array}
\end{equation}

\begin{equation}
\hspace{2 mm} V_Z= V_r\, \sin b + V_b\, \cos b +V_{Z0}, \label{vz}
\end{equation}

\noindent where the angle $\alpha$ is:

\begin{equation}
\alpha= \ell +\theta- \pi/2
\end{equation}

\noindent and $V_r$ is the line-of-sight velocity. The stellar
velocities along the Galactic longitude $V_l$ and latitude $V_b$ are
determined from the relations:

\begin{equation}
V_l= 4.74 \, \mu_l \,r , \label{vl}
\end{equation}
\begin{equation}
V_b= 4.74 \, \mu_b \, r, \label{vb}
\end{equation}

\noindent where $\mu_l$ and $\mu_b$ are {\it Gaia}  proper motions
along the Galactic longitude and latitude, respectively.  The factor
$4.74\times r$~(kpc) transforms  units of mas yr$^{-1}$ into km
s$^{-1}$.

\begin{figure*}
\resizebox{\hsize}{!}{\includegraphics{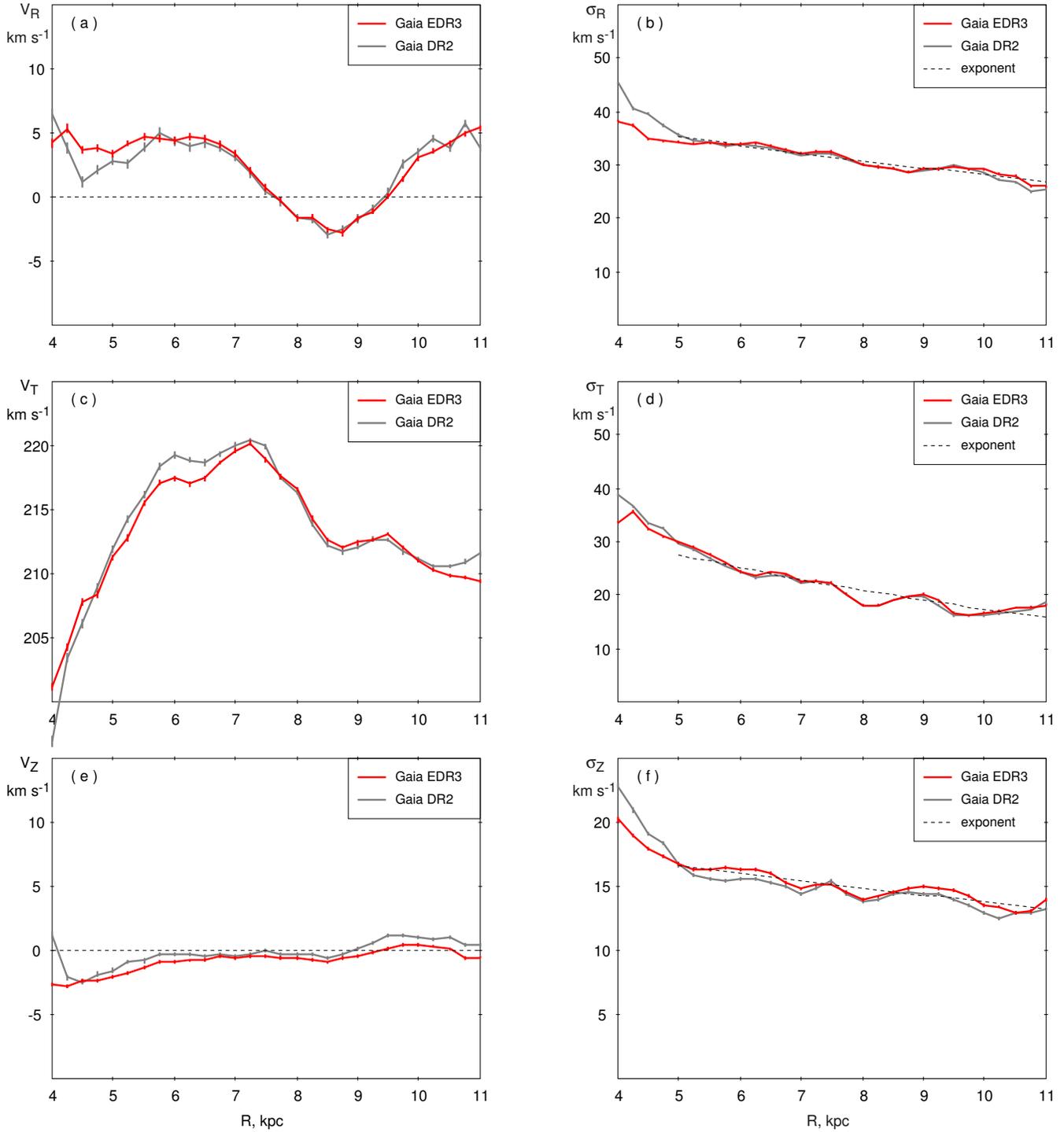}} \caption{Variations
in the velocity components, $V_R$, $V_T$ and $V_Z$ (left panel), and
in the velocity dispersions, $\sigma_R$, $\sigma_T$ and $\sigma_Z$
(right panel), along the Galactocentric distance, $R$,  derived from
the {\it Gaia} EDR3 and {\it Gaia} DR2 data. The median velocities
are calculated  in 250-pc wide bins. The vertical lines on the curves
demonstrate the random errors in the determination of the velocities
and velocity dispersions. We can see (a) that the radial velocity
$V_R$ reaches a maximum of +5 km s$^{-1}$ at the distance of $R\sim
6$ kpc and a minimum of $V_R\sim -3$ km s$^{-1}$ at $R\sim 8.5$ kpc.
Note also that the $V_R$-profile derived from the {\it Gaia} DR3 data
is flatter in the distance interval 4--6 kpc than the profile derived
from the {\it Gaia} DR2 catalogue. (c) The profile of the azimuthal
velocity $V_T$ demonstrates a sharp drop by $\Delta V_T=7$ km
s$^{-1}$ in the distance interval $R=7.0$--8.5 kpc. Noticeably lower
velocities $V_T$ in the distance range 4.0--5.5 kpc can be due to the
lack of thin-disc stars in this region. (e) The vertical velocity
$V_Z$  remains within $\pm 1$ km s$^{-1}$ in the distance interval
6.0--9.0 kpc but exhibits a  velocity bend at the larger distances
from the Sun, which can be due to the wrong distance scale. (b, d, f)
The variations in the velocity dispersions, $\sigma_R$, $\sigma_T$
and $\sigma_Z$,  with the distance $R$ were approximated by the
exponential law (the dashed lines).} \label{obs_prof}
\end{figure*}

Fig.~\ref{obs_prof} (left panel) shows the variations in the median
velocities, $V_R$, $V_T$ and $V_Z$,  with the Galactocentric
distance, $R$, derived from the {\it Gaia} EDR3 and {\it Gaia} DR2
data. The median velocities are calculated  in $\Delta R=250$-pc wide
bins.   The  average errors in the determination of the median {\it
Gaia} EDR3  velocities  $V_R$, $V_T$ and $V_Z$ in bins at the
interval $R=5$--10 kpc are  0.16, 0.11 and 0.07 km s$^{-1}$,
respectively.

Fig.~\ref{obs_prof}(a)  shows that the radial velocity $V_R$ reaches
a  maximum of +5 km s$^{-1}$ at the distance of $R\sim 6$ kpc and a
minimum of $V_R\sim -3$ km s$^{-1}$ at $R\sim 8.5$ kpc. We can also
see that the $V_R$-profile derived from the {\it Gaia} DR3 data is
flatter  in the distance interval 4--6 kpc than the profile derived
from the {\it Gaia} DR2 data.

Fig.~\ref{obs_prof}(c) shows a sharp drop of the azimuthal velocity,
$V_T$,  by $\Delta V_T=7$ km s$^{-1}$ in the distance interval
$R=7.0$--8.5 kpc. We can also see   noticeably lower  velocities
$V_T$ in the distance interval $R=4.0$--5.5 kpc, which can be due to
the lack of thin-disc stars in this region. The extinction in the
middle of the Galactic plane grows very rapidly in the direction
toward the Galactic center \citep{neckel1980, marshall2006,
melnik2016}, so our sample of {\it Gaia} stars at the distances of
$R=4.0$--5.5 kpc can contain a larger fraction of stars associated
with the thick disc and halo than in other regions.  The sources for
which a line-of-sight velocity is  listed in the Gaia DR2 and EDR3
catalogues mostly have the magnitude brighter than $G=13^m$. So the
fraction of thin and thick disk stars as well as the proportion of
young versus old stars must depend on the place in the Galactic disc
\citep{brown2018, sartoretti2018, katz2018, katz2019}.

Fig.~\ref{obs_prof}(e) shows that the vertical velocity $V_Z$  does
not exceed $\pm 1$ km s$^{-1}$ in the distance interval 6.0--9.0 kpc
but demonstrates the velocity bend at greater heliocentric distances:
$V_Z$ is negative ($V_Z=-2.0\pm0.1$ km s$^{-1}$) in the distance
interval of $R=4.0$--5.0 kpc and  positive ($V_Z=+0.5\pm0.1$ km
s$^{-1}$)   at  $R=10.0$ kpc. Generally, the $V_Z$-velocity bend can
be due to  two circumstances: wrong distance scale at large
heliocentric distances, $r>2$ kpc, plus some ripple of the Galactic
disc.  The preponderance of objects at negative or positive Galactic
latitudes (the ripple) produces some excess in the average value of
the term  $V_r\, \sin b$ (Eq.~\ref{vz}), which can be uncompensated
by the term $V_b\, \cos b$ due to the distance-scale errors, because
$V_b$ is directly proportional to the heliocentric distance $r$
\citep[Eq.~\ref{vz} and ~\ref{vb}, see also discussion
in][]{melnik2020}.

Fig.~\ref{obs_prof} (right panel) shows  the distributions of the
velocity dispersions in the  radial, azimuthal and vertical
directions,  $\sigma_R$, $\sigma_T$ and $\sigma_Z$, along the
Galactocentric distance, $R$, derived from the {\it Gaia} EDR3 and
{\it Gaia} DR2 data. The median velocity dispersion in each bin was
determined as  half of the central velocity interval including 68 per
cent of objects. We approximated the variations in the velocity
dispersions with the distance $R$  by the exponential law and
obtained the following dependencies for the {\it Gaia} EDR3 data in
the distance interval $R=5$--11 kpc:

%------------------   Table 1   -----------------------------------------------
\begin{table}  \caption{Parameters derived from EDR3 data}
  \begin{tabular}{cc}
%  \hline
  \\[-7pt]\hline\\[-7pt]
  $C_R=44.0\pm1.0$ km s$^{-1}$ & $S_R=22.3\pm1.4$ kpc  \\
  $C_T=43.8\pm2.8$ km s$^{-1}$ & $S_T=10.8\pm0.9$ kpc  \\
  $C_Z=20.0\pm0.6$ km s$^{-1}$ & $S_Z=26.9\pm2.6$ kpc  \\
  \\[-7pt]\hline\\[-7pt]
%\hline
\end{tabular}
\label{tab_par}
\end{table}
%------------------------------------------------------------------------------

\begin{equation}
\sigma_R = C_R \; \exp^{-\frac{R}{S_R}}  , \label{sr_obs}
\end{equation}
\begin{equation}
\sigma_T = C_T \; \exp^{-\frac{R}{S_T}}  ,
\end{equation}
\begin{equation}
\sigma_Z = C_Z \; \exp^{-\frac{R}{S_Z}}  ,
\end{equation}

\noindent where the values of the parameters and their errors are
listed in Table~\ref{tab_par}. Note that our estimate of the scale
length $S_R=22.3\pm1.4$ kpc is consistent with the value calculated
by \citet{eilers2019}, $S_R=21$ kpc. The characteristic scales
derived from the {\it Gaia} DR2 catalogue ($S_R=19.9\pm1.1$ kpc,
$S_T=10.7\pm0.9$ kpc and $S_T=26.4\pm2.4$ kpc)  agree with the scale
lengths calculated for the {\it Gaia} EDR3 data. Such a large
difference between the values of $S_R$ and $S_T$  can be due to
systematic effects. The space distribution of stars in our sample
(Fig.~\ref{gaia_distrib}a) suggests that  the dispersion of radial
velocities, $\sigma_R$, is mainly determined by line-of-sight
velocities while those of azimuthal  velocities, $\sigma_T$, is
mostly determined by proper motions and parallaxes, which are subject
to systematic effects to a greater extent.

The dispersion of radial velocities $\sigma_R$ at the solar distance
amounts to $\sigma_R=32.5$ km s$^{-1}$ while the value derived from
the smoothed distribution (Eq.~\ref{sr_obs}) is  $\sigma_R=31.5$ km
s$^{-1}$.

The root-mean-square deviations between the   median velocities  in
bins derived from the {\it Gaia} DR2 data and from  {\it Gaia} EDR3
data in the distance interval 5.5--9.5 kpc amount to  0.4, 0.9 and
0.5 km s$^{-1}$ for the components $V_R$, $V_T$ and $V_Z$,
respectively.

%-----------------------    Figure 1  -----------------------------------------
%------------------------------------------------------------------------------

%-----------------------    Figure 2  -----------------------------------------
%------------------------------------------------------------------------------

%-----------------------    Figure 3  -----------------------------------------
\begin{figure*}
\resizebox{\hsize}{!}{\includegraphics{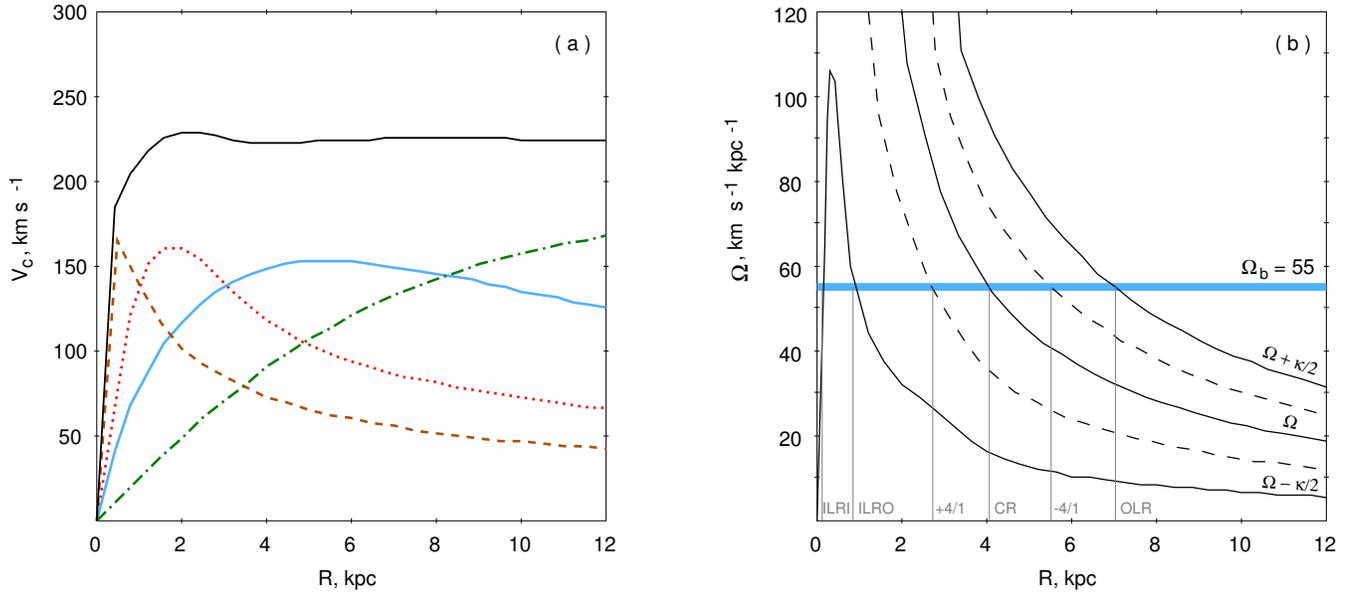}} \caption{ (a) Model
rotation curve. The total rotation curve  (the black curve) and the
contribution of the bulge, bar, disc and halo (the brown, red, blue
and green curves, respectively) to the rotation curve. (b)
Dependencies  of the angular velocities $\Omega$, $\Omega\pm\kappa/2$
(the solid curves) and $\Omega \pm \kappa/4$ (the dashed curves) on
the distance $R$. The horizontal straight line (coloured  in blue)
indicates the angular velocity of the bar and its intersections with
the curves of angular velocities mark the locations of the
resonances.} \label{rot_curve}
\end{figure*}
%------------------------------------------------------------------------------

%-----------------------    Figure 4  -----------------------------------------
\begin{figure*}
\resizebox{\hsize}{!}{\includegraphics{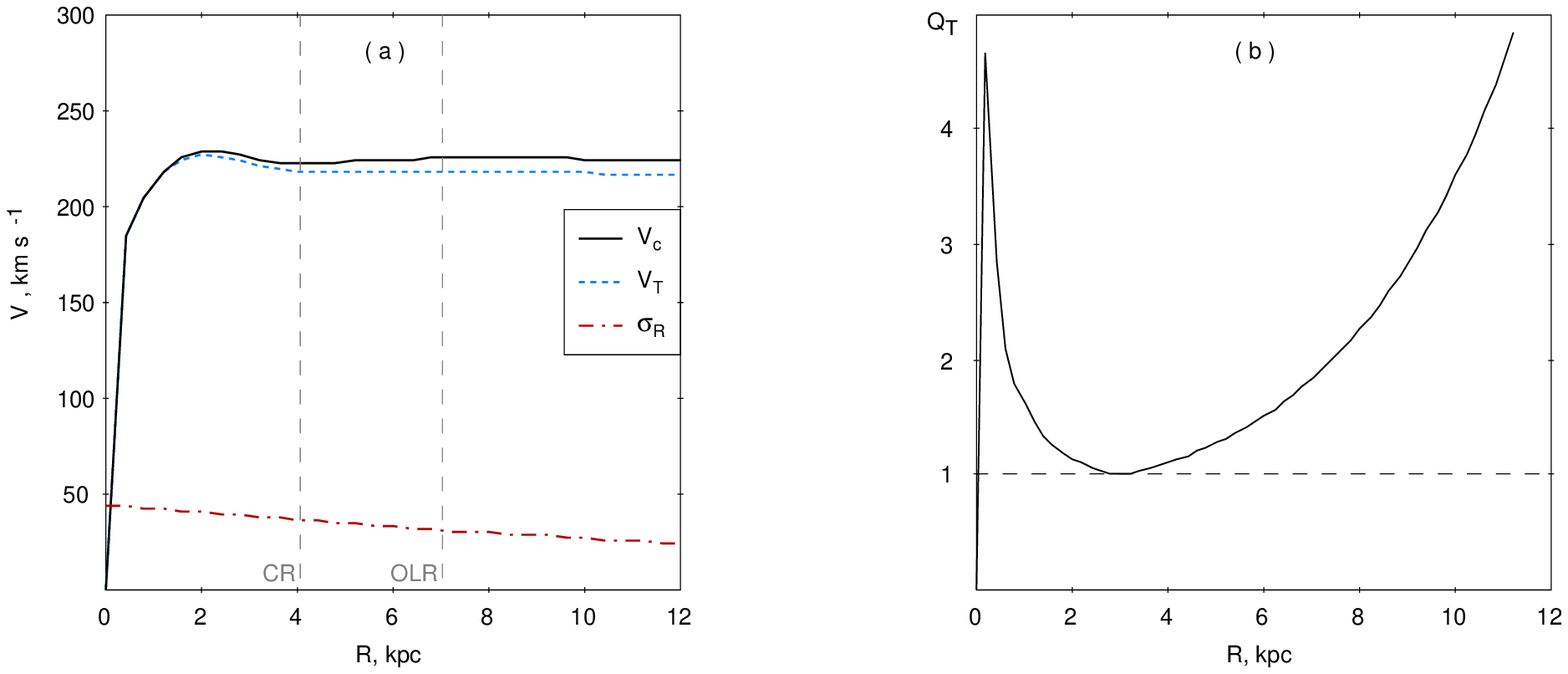}} \caption{Initial
velocity distribution of model particles. (a) Velocities of the
rotation curve  $V_c$ (solid line), the average azimuthal velocities
$\overline{V_T}$ (dashed line) and the initial velocity dispersion
$\sigma_R$ (dash-and-dot line) of model particles. (b) Dependence of
the Toomre disc stability parameter $Q_T$ on Galactocentric distance
$R$ built for the initial velocity distribution.} \label{az_vel}
\end{figure*}
%------------------------------------------------------------------------------

\section{Model}

\subsection{General remarks}

In the previous section we showed that there are no conspicuous
systematic motions in the direction perpendicular to the Galactic
disc in  the vicinity of $|R-R_0|<1.5$ kpc from the Sun. So the
motions in the Galactic plane and in the vertical direction can be
thought to be independent, which gives us a possibility to consider a
2D model of the Galaxy.

We built  a  model of the Galaxy that includes the bulge, bar,
exponential disc and halo. The bar is modelled as a Ferrers ellipsoid
with the volume-density distribution $\rho$  defined as follows:

\begin{equation}
\rho = \left\{ \begin{array}{l}
\rho_0 (1-\xi^2)^n, \ \xi \leq 1\\
\\
0, \ \xi > 1,\\
\end{array} \right.
\label{kalnajs}
\end{equation}

\noindent  where  $\rho_0$ is the central density, $\xi$  equals
$\xi^2=x^2/a^2+y^2/b^2$ but $a$ and $b$ are the lengths of the major
and minor semi-axes of the bar, respectively.  The exponent $n$ is
adopted to be  $n=2$. The mass of the bar $M_b$ and  the central
density $\rho_0$ are  related by the expression:

\begin{equation}
M_b =\rho_0 \,\frac{32\pi a b^2}{105} \label{m_bar}
\end{equation}

\noindent \citep{freeman1970,athanassoula1983,pfenniger1984,
binney2008, sellwood1993}.

The main model requirement is  that the model must reproduce the
observed profiles of the radial, $V_R$, and azimuthal, $V_T$,
velocities as well as the profiles of the velocity dispersions,
$\sigma_R$ and $\sigma_T$ (Fig.~\ref{obs_prof}). In addition,  the
model rotation curve must be flat on the Galactic periphery  and have
the angular velocity  of $\Omega_0=30$ km s$^{-1}$ kpc$^{-1}$ at the
solar distance, which agrees with observations and corresponds to the
solar azimuthal velocity $V_{T0}=\Omega_0 R_0+12$ km s$^{-1}$ used
for calculations of stellar velocities with respect to the Galactic
center (section ~\ref{nd}).

We adopted the simulation time to be  $T=3$ Gyr. The model includes
$N=10^7$ particles that simulate the motion of stars in the Galactic
disc. This number is enough to produce the velocity profiles with an
accuracy comparable to the observational one.

Fig.~\ref{rot_curve}(a) shows the contribution of the bulge, bar,
disc and halo to the total rotation curve. The total rotation curve
is practically flat on the periphery and corresponds to the angular
velocity of the disc rotation   at the solar distance equal to
$\Omega_0=30.0$ km s$^{-1}$ kpc$^{-1}$.

Fig.~\ref{rot_curve}(b) shows the dependencies  of the angular
velocities $\Omega$, $\Omega\pm\kappa/2$ and $\Omega \pm \kappa/4$ on
Galactocentric distance $R$. The horizontal line indicates the
angular velocity of the bar, and its intersections with the curves of
angular velocities mark the locations of the resonances.

Table~\ref{general} lists the model parameters of the bar, bulge,
disc and halo. The semi-major and semi-minor axes  of the bar are
adopted to be $a=3.5$ and $b=1.35$ kpc, respectively. The mass of the
bar is 1.2 10$^{10}$ M$_\odot$. The angular velocity of the bar is
taken to be $\Omega_b=55$ km s$^{-1}$ kpc$^{-1}$, which corresponds
to the location of the corotation radius $(\Omega=\Omega_b)$ and the
OLR of the bar at the distances of $R_{CR}=4.04$ kpc and
$R_{OLR}=7.00$ kpc, respectively. The model also includes two Inner
Lindblad Resonances (ILRs) located at the distances of
$R_{ILRI}=0.14$ and $R_{ILRO}=0.87$ kpc.

Non-axisymmetric perturbations of the bar increase slowly approaching
the full strength by $T_{gr}=447$ Myr, which is equal to four bar
rotation periods. However, the $m=0$ component of the bar is included
in the model from the beginning.   During the period of the bar
growth, $0<t<T_{gr}$, radial and azimuthal forces created by the bar
were multiplied by the factor $k_g$ which increases linearly from 0
to 1. We also introduce the additional force, $f_a$, which on the
contrary decreases linearly with time:

\begin{equation}
 \left \{
 \begin{array}{lc}
 k_g=t/T_{gr} &  \\
 & 0<t<T_{gr},\\
 f_a=(1-k_g)\,<f_b> &\\
 \end{array}
 \right.\\
  \label{p}
\end{equation}

\noindent where $<f_b>$ is the radial force created by the bar
averaged over the azimuthal angle $\theta \in [0,2\pi]$. The force
$<f_b>$ depends on the radius, $R$, so its values were  tabulated
before the simulation and then were calculated for a position of a
particle  by linear interpolation. This procedure ensures a constant
average radial force at each radius during the bar-growth period. The
$m=0$ component of the bar can be interpreted as a pre-existent
disc-like bulge \citep{athanassoula2005}.

The  $q_t(R)$ quantity is often considered as the characteristic of
perturbations created by a bar. It is calculated  as the maximal
ratio of the acceleration in the azimuthal direction to the average
acceleration in the radial direction at a certain radius:

\begin{equation}
 q_t(R)= \textrm{max}(\frac{\max(|F_T|)}{<|F_R|>}).
\label{def_qt}
\end{equation}

\noindent The strength of the bar, $Q_b$, is determined as the
maximal value of $q_t$ along the radius \citep{sanders1980,
combes1981, athanassoula1983}. In the present model  the strength of
the bar amounts to $Q_b=0.3142$, which is a typical value for
galaxies with strong bars \citep{block2001, buta2004,
diaz-garcia2016}.

The model of the Galaxy includes an exponential disc with a mass of
$M_d=3.25$ 10$^{10}$M$_\odot$ and a characteristic  scale of
$R_d=2.5$ kpc \citep{bland2016}. The total mass of the model disc and
bar is $M_d+M_b=4.45 \times 10^{10}$M$_\odot$, which agrees with
other estimates of the Galactic-disc mass lying in the range
3.5--5.0$\times 10^{10}$M$_\odot$ \citep{shen2010, fujii2019}. The
surface density of the disc is determined by the relation:

\begin{equation}
\Sigma=\Sigma_0 \, \exp^{-\frac{R}{R_D}}, \label{disc}
\end{equation}

\noindent where $\Sigma_0$ is the density of the disc at the Galactic
center. Here we consider only the thin disc because the ratio of the
surface densities of the thick  and thin  discs are supposed to be
only 12 per cent at the solar distance \citep[as a
review][]{bland2016}.

The  classical bulge is  modelled by a Plummer sphere with the mass
of $M_{bg}=5\times 10^{9}$ \citep{dehnen1998, nataf2017, fujii2019}.
The halo is modelled by an isothermal sphere. More detailed
description of the construction method for each Galactic subsystem is
given in \citet{melnik2019}.

%--------------------------- Table 2 ------------------------------------------
\begin{table}  \caption{Model parameters}
  \begin{tabular}{ll}
%  \hline
  \\[-7pt]\hline\\[-7pt]
Simulation time &  $T=3$  Gyr  \\
Step of integration     &  $\Delta t=0.01$ Myr \\
Number of particles     &  $N=10^7$\\
\hline
Bulge & $R_{bg}=0.30$  kpc    \\
      & $M_{bg}=5$ 10$^{9}$M$_\odot$  \\
\hline
Bar   & $a=3.5$  and $b=1.35$ kpc \\
      & $M_b=1.20$ 10$^{10}$M$_\odot$ \\
      & $\Omega_b=55.0$ km s$^{-1}$ kpc$^{-1}$  \\
      & $T_{gr}=447$ Myr  \\
\hline
Disc  & exponential, $R_d=2.5$ kpc \\
      & $M_d=3.25$ 10$^{10}$M$_\odot$ \\
\hline
Halo &  $R_h=8$  kpc   \\
     &  $V_{max}=201.4$  km s$^{-1}$  \\
  \\[-7pt]\hline\\[-7pt]
%\hline
\end{tabular}
\label{general}
\end{table}
%------------------------------------------------------------------------------

%-----------------------    Figure 5  -----------------------------------------
\begin{figure}
\resizebox{\hsize}{!}{\includegraphics{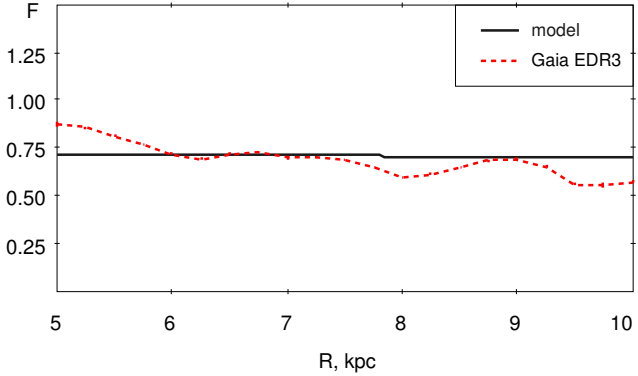}} \caption{Dependence
of the ratio of the observed velocity dispersions,
$\sigma_T/\sigma_R$, derived from the {\it Gaia} EDR3 data (the red
dashed line) and the model value $F=\kappa/(2\Omega)$ (the black
line) on  Galactocentric distance $R$. The value of $F$ determines
the ratio of the azimuthal and radial velocity dispersions adopted
for the initial distribution of model particles.} \label{dis_rat}
\end{figure}
%------------------------------------------------------------------------------

\subsection{The initial velocity distribution}

We suppose that 3 Gyr ago  the dependence of the radial velocity
dispersion $\sigma_R$ on the distance $R$ in the  $R=5$--11 kpc
interval was similar to what is observed now:

\begin{equation}
\sigma_R = C^*_R \exp^{-\frac{R}{S_R}}, \label{srm}
\end{equation}

\noindent where $S_R$ is adopted to be $S_R=20$ kpc
(Eq.~\ref{sr_obs}).

We want to make sure that the model disc is stable against
axisymmetric perturbations at the initial time instant. So the value
of the constant  $C^*_R$ is determined so as to ensure the Toomre
stability parameter would be $Q_T \ge 1$ throughout the disc
\citep{toomre1964}. In our model,  $Q_T$ achieves a minimum at the
distance of $R=3.1$ kpc, so the constant $C^*_R$ is determined from
the equation:

\begin{equation}
Q_T=\frac{\sigma_R \,\kappa }{3.36 \, G \,\Sigma} =1
\end{equation}

\noindent at the distance of $R=3.1$ kpc (Fig.~\ref{az_vel}b). This
normalization yields the initial radial-velocity dispersion at the
solar distance of $\sigma_R=30.5$ km s$^{-1}$, which is very close to
the value of $\sigma_R=31.5$ km s$^{-1}$ derived from the smooth
distribution of the Gaia velocity dispersions (Eq.~\ref{sr_obs}).
After the model bar started growing, the velocity dispersion inside
the corotation radius ($R_{CR}=4.0$ kpc) increased rapidly but we
want the disc to be stable at the initial moment as well (see also
section ~\ref{vd}).

%The average radial velocity of model particles is zero at the initial
%moment, $\overline{V_R}=0$, and the initial velocity dispersion,
%$\sigma_R$, is determined by the equation Eq.~\ref{srm}.

Any subsystem of disc stars with a non-zero velocity dispersion
rotates with a smaller average velocity than the velocity of the
rotation curve \citep[for example,][]{binney2008}. The relation
between the average azimuthal velocity $\overline{V_T}$ and the
velocity of the rotation curve $V_c$ is determined by the Jeans
equation:

\begin{equation}
\frac{\partial (\rho \overline{V^2_R})}{\partial R} +
\frac{\partial(\rho\overline{V_R V_z})}{\partial z} + \rho
\;(\frac{\overline{V^2_R} - \overline{V^2_T}}{R} +
\frac{\partial\Phi}{\partial R})  = 0, \label{jeans1}
\end{equation}

\noindent where $\rho$ is the volume density of disc stars and $\Phi$
is the Galactic potential which is related to the   velocity of the
rotation curve by the following expression:

\begin{equation}
V^2_c=R\, \frac{\partial\Phi}{\partial R}. \label{vc}
\end{equation}

\noindent Assuming that the motions along coordinates  $R$ and $z$
are independent, we can neglect the second term in the left part of
Eq.~\ref{jeans1}. If we suppose that the distribution law along the
coordinate $z$ does not depend on the distance $R$, then we can
substitute the volume density  $\rho$ for the surface density
$\Sigma$. We assume that systematic motions are absent at the initial
moment  and can make following substitutions:

\begin{equation}
\overline{V^2_R}=\sigma^2_R, \label{vr}
\end{equation}
\begin{equation}
\overline{V^2_T}=\overline{V_T}^2 + \sigma^2_T, \label{vt}
\end{equation}

\noindent which allows us to rewrite the Jeans equation in the
following way:

\begin{equation}
\overline{V_T}^2 = V^2_c + \sigma_R^2 -\sigma_T^2 +\frac{R}{\Sigma}
\frac{\partial (\Sigma \sigma_R^2)}{\partial R}.
\end{equation}

\noindent Assuming the epicyclic approximation for the initial
velocity distribution, we can write the following relation between
the velocity dispersions along the radial and azimuthal directions:

\begin{equation}
\frac{\sigma_T}{\sigma_R} = \frac{\kappa}{2\,\Omega}. \label{stsr}
\end{equation}

\noindent We now use Eqs.~\ref{sr_obs},~\ref{disc} and~\ref{stsr}  to
obtain:

\begin{equation}
\overline{V_T}^2 = V_c^2 + \sigma_R^2 (1 -\frac{\kappa^2}{4\Omega^2}
-\frac{R}{R_D}-\frac{2R}{S_R}), \label{jeans2}
\end{equation}

\noindent where  $\overline{V_T}$ is the average azimuthal velocity
of model particles at the initial time instant.

Fig.~\ref{az_vel}(a) shows the distribution of the average azimuthal
velocity, $\overline{V_T}$, of model particles  along Galactocentric
distance, $R$, at the initial instant. Also shown are the velocity of
the rotation curve, $V_c$, and  the radial-velocity dispersion,
$\sigma_R$, at the initial moment. We can see that the velocity
$\overline{V_T}$ is always slightly lower than the velocity of the
rotation curve and the difference $V_c-\overline{V_T}$ increases with
the increasing distance $R$.

Fig.~\ref{dis_rat} shows the dependence of the ratio of the observed
velocity dispersions of disc stars, $\sigma_T/\sigma_R$,  and the
model value $F=\kappa/(2\Omega)$ on Galactocentric distance $R$. The
value of $F$ determines the ratio of the azimuthal and radial
velocity dispersions adopted for the initial distribution of model
particles. We can see that  $F$ is almost  constant and varies in the
range $F=0.70$--0.72 in the distance interval  5--10 kpc, which is
quite expected for model with a flat rotation curve for which $F$
must be equal to $F=1/\sqrt{2}$.  The formal errors in the
determination of the observed ratio $\sigma_T/\sigma_R$ are less than
0.01. The model value $F=\kappa/(2\Omega)$ is derived from the model
potential so it is absolutely accurate. We can see that the {\it
Gaia} EDR3 ratio of the velocity dispersions, $\sigma_T/\sigma_R$,
decreases with increasing distance $R$, which can be due to the rapid
change of the velocity dispersion $\sigma_T$ with $R$
(Table~\ref{tab_par}). However, in the solar neighborhood of
$|R-R_0|<1.5$ kpc, the observed ratio $\sigma_T/\sigma_R$  lies in
the range 0.60--0.72. So here the model and observed values of the
velocity dispersions  are consistent to within 15\%.

\section{Results}

\subsection{Formation of the rings}

\begin{figure*}
\resizebox{\hsize}{!}{\includegraphics{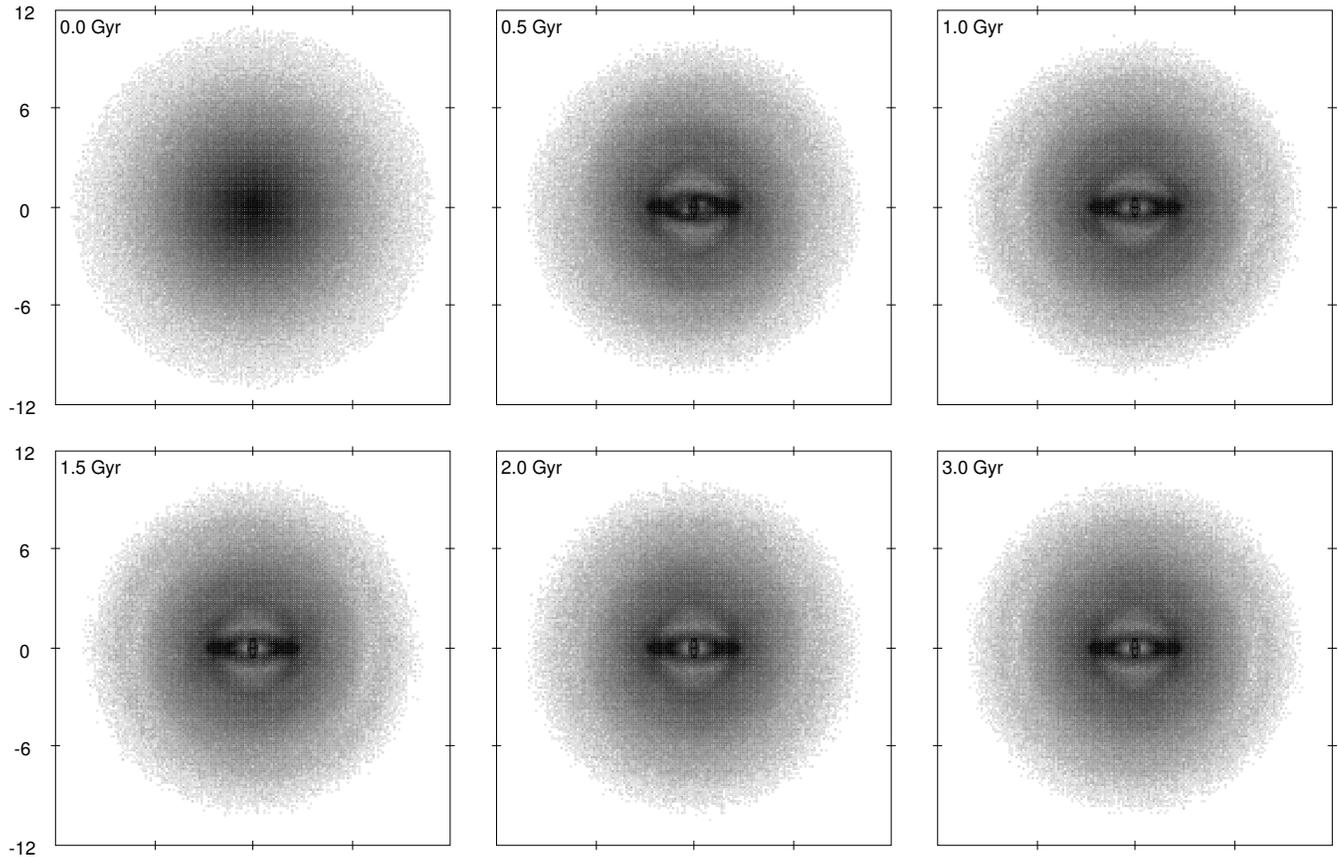}}
\caption{Distribution of  the surface density of model particles in
the Galactic disc at different time moments. The size of the frames
is $24\times 24$ kpc. The major axis of the bar is oriented
horizontally.  Two closest shades of gray correspond to a density
change by a factor of 0.8. The model galaxy rotates
counterclockwise.} \label{distrib}
\end{figure*}

The formation of   elliptical resonance rings requires some time for
the epicyclic motions of stars  to adjust in accordance with the
rotation of the bar. Fig.~\ref{distrib} shows  the distribution of
the surface density  of model particles in the Galactic disc at
different time instants. The time $t=0$ corresponds to the instant
when the bar is turned on. By the time $T_{gr}=0.45$ Gyr the bar
gains the full strength. One can see that the outer rings, $R_1$ and
$R_2$, are already formed by the time $t=1$ Gyr. The ring $R_1$ is
located closer to the Galactic center and is stretched perpendicular
to the bar so that the bar and ring $R_1$ are reminiscent of the
Greek letter '$\theta$'. The ring $R_2$ is located further away from
the Galactic center and is stretched parallel to the bar. Besides the
outer rings, $R_1$ and $R_2$, the Galactic disc produces the inner
($r$) and nuclear ($n$) resonance rings.

\begin{figure*}
\resizebox{\hsize}{!}{\includegraphics{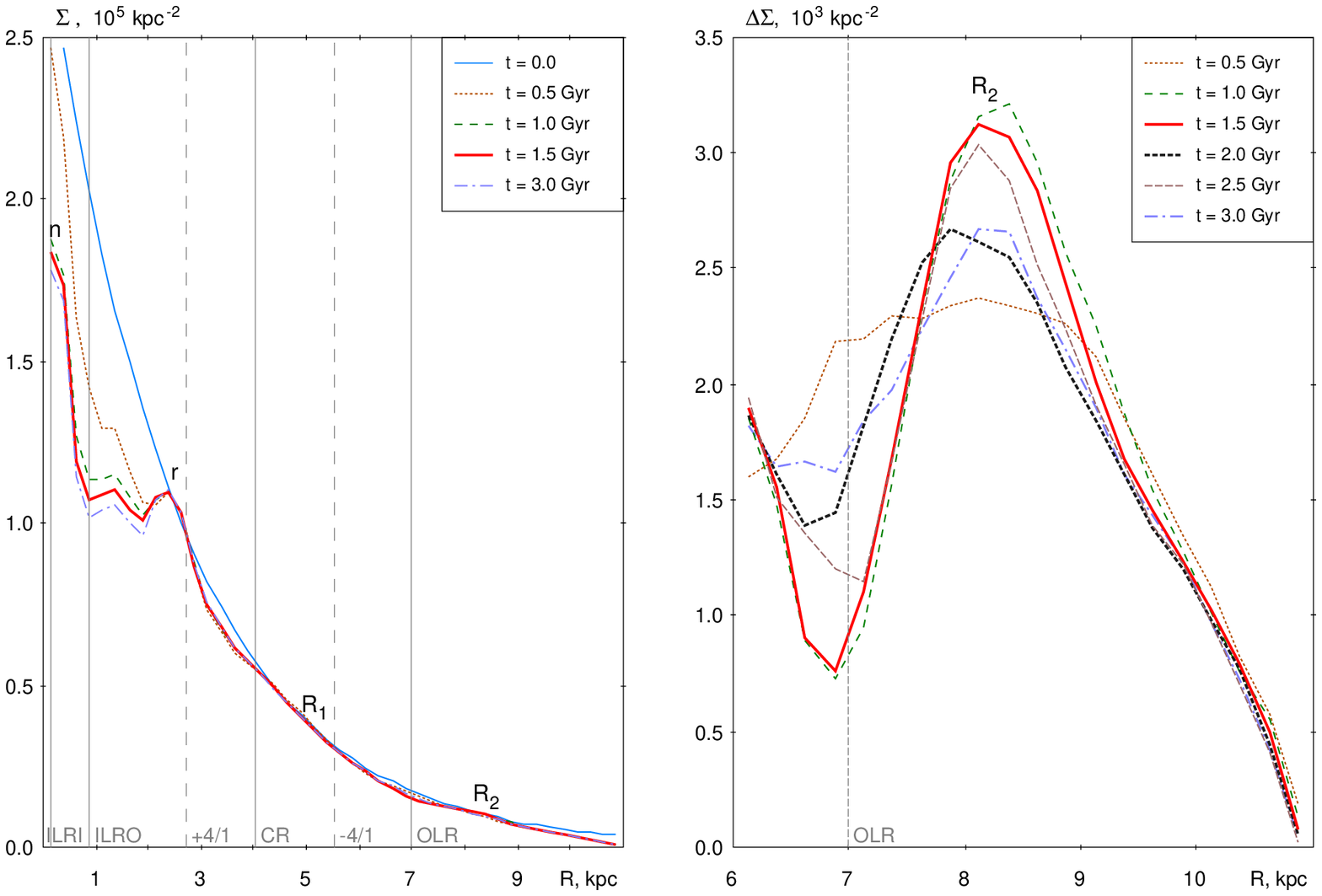}} \caption{(a)
Distribution of the surface density, $\Sigma$, of model particles
along Galactocentric distance $R$ at different time instants. The
locations of the resonance rings are marked by the letters: $n$
(nuclear ring), $r$ (inner ring), $R_1$ and $R_2$ (outer rings). (b)
Distribution of the  relative surface density,
$\Delta\Sigma=\Sigma-\Sigma_0+C$, of model particles near the OLR,
where $\Sigma_0$ is the initial exponential density distribution but
the constant $C=3.2$ 10$^3$ kpc$^{-2}$ is added to avoid dealing with
negative values. We can see some variations in the width and density
of the outer ring $R_2$: the ring $R_2$ has a flatter profile at the
times $t=2.0$ and 3.0 Gyr than at the times $t=1.0$ and 1.5 Gyr.}
\label{den_distrib}
\end{figure*}

Fig.~\ref{den_distrib}(a) shows the distribution of the surface
density, $\Sigma$, of model particles along the distance $R$ at
different time moments. Local density minima  between the following
pairs of rings: $n$ and $r$, $r$ and $R_1$, $R_1$ and $R_2$, are
immediately apparent. Interestingly, the location on the OLR
precisely coincides with the density minimum.

Fig.~\ref{den_distrib}(b) shows the distribution of the  relative
surface density, $\Delta\Sigma=\Sigma-\Sigma_0+C$, of model particles
near the OLR.  The function $\Sigma_0$ describes the exponential
density distribution at the start of the simulation but the constant
$C=3.2$ 10$^3$ kpc$^{-2}$ is added to avoid dealing with negative
values. We can see some variations in the width and density of the
outer ring $R_2$: at the times $t=2.0$ and 3.0 Gyr the ring $R_2$ has
larger width and  smaller density than at the times $t=1.5$ and 1.0
Gyr, which is mainly due to the shift of its inner boundary.
Probably, here we see the periodically strengthening and weakening
ring $R_2$.

%------------------------------------------------------------------------------
%-----------------------    Figure 6  -----------------------------------------
%------------------------------------------------------------------------------
%------------------------------------------------------------------------------
%-----------------------    Figure 7  -----------------------------------------
%------------------------------------------------------------------------------
%------------------------------------------------------------------------------

\subsection{Comparison between the model and observed velocity profiles}

\begin{figure*}
\resizebox{16 cm}{!}{\includegraphics{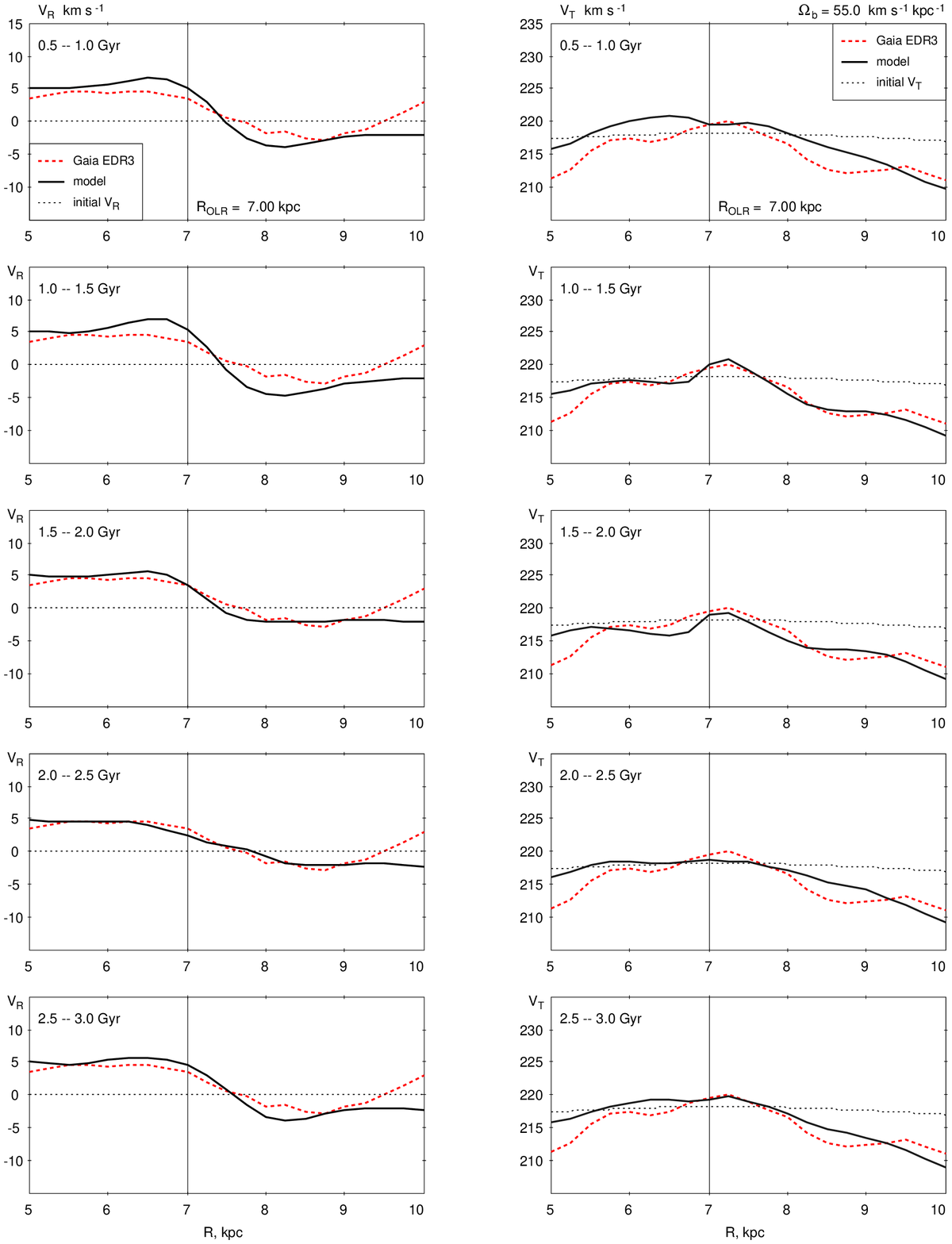}} \caption{Model and
observed dependencies of the median radial and azimuthal velocities,
$V_R$ and $V_T$, of disc stars lying in the sector
$|\theta|<15^\circ$ on Galactocentic distance $R$. The model velocity
profiles (the black curves) are obtained for the angular velocity and
the position angle of the bar equal to $\Omega_b=55$ km s$^{-1}$
kpc$^{-1}$ and $\theta_b=45^\circ$, respectively,  and averaged over
the time periods of 0.5 Gyr. The observed velocity profiles (the red
dashed curves) are derived from the {\it Gaia} EDR3 data.  The errors
in determination of both {\it Gaia} EDR3 and model average velocities
in bins are smaller than the line width. The dashed gray lines
indicate the initial distribution of the velocities $V_R$ and $V_T$.
The boundaries of time periods are shown at each frame. The vertical
lines indicate the location of the OLR.} \label{prof_55}
\end{figure*}

Fig.~\ref{prof_55} shows the model and observed  dependencies of the
radial, $V_R$, and azimuthal,  $V_T$, velocities of disc stars lying
in the sector $|\theta|<15^\circ$ on Galactocentric distance $R$. The
model dependencies are obtained for the angular velocity and the
position angle of the bar equal to  $\Omega_b=55$ km s$^{-1}$
kpc$^{-1}$ and $\theta_b=45^\circ$, respectively.  The position angle
of the bar close to $\theta_b=45^\circ$ is found in many studies
\citep[][and other papers]{hammersley2000, benjamin2005,
cabrera-lavers2007, melnikrautiainen2009, gonzalez2012, pettitt2014}.
The median velocities of model particles were calculated in $\Delta
R=250$-pc wide bins. Model velocity profiles were averaged over 0.5
Gyr time periods with  a step of $\Delta t=0.01$ Gyr.  The random
errors in the determination of the average model velocities  in bins
are smaller than 0.1 km s$^{-1}$. The observed velocity profiles were
derived from the {\it Gaia} EDR3 data (section 2).

Fig.~\ref{prof_55} (left panel) shows the model and observed profiles
of the radial velocity  $V_R$. The  model $V_R$-profiles demonstrate
a plateau in the distance interval $R=5.0$--6.0 kpc, a smooth fall
from the velocity $V_R=+5$ to $-3$ km s$^{-1}$ over the interval
$R=6.0$--8.5 kpc and a slow rise or a plateau with a constant
negative velocity $V_R$ in the interval $R=8.5$--10.0 kpc. We can see
that the model and observed $V_R$-profiles match well in the distance
interval $R=6$--9 kpc. Note that the model profiles of the velocity
$V_R$ obtained during the time periods $t=0.5$--1.0 and 1.0--1.5 Gyr
are steeper than the profiles calculated for the periods 1.5--2.0,
2.0--2.5 and 2.5--3.0 Gyr, which can result from a small increase in
the velocity dispersion due to the tuning of the epicyclic motions
near the OLR (i.e. to the formation of the outer rings). Generally,
the model and observed $V_R$-profiles match better during the time
periods 1.5--2.0, 2.0--2.5 and 2.5--3.0 Gyr than at the times
0.5--1.0 and 1.0--1.5 Gyr.

Fig.~\ref{prof_55} (right panel) shows the model and observed
profiles of the azimuthal velocity $V_T$. The  model velocity $V_T$
is almost constant, $V_T=216$--218 km s$^{-1}$, at the start of the
simulation (the slightly curved dashed line). The velocity $V_T$
drops by 5 km s$^{-1}$ at the distance of $R=8.5$ kpc by the time
$t=1.0$ Gyr. We can see that the model and observed $V_T$-profiles
agree well over the distance interval $R=6.0$--10.0 kpc at the time
periods 1.0--1.5 and 1.5--2.0 Gyr. The model can reproduce the sharp
drop of the velocities $V_T$ in the distance interval $R=7.5$--8.5
kpc just at the times 1.0--1.5 and 1.5--2.0 Gyr, but later the model
velocity $V_T$ demonstrates a smoother fall.

The study of the influence of the observational errors and selection
effects onto the distributions of the velocities $V_R$ and $V_T$
along the distance $R$ is presented in Appendix.

%-----------------------    Figure 8  -----------------------------------------
%------------------------------------------------------------------------------

\subsection{Comparison of the velocity profiles during different time periods}\label{time}

We calculated the statistics $\chi^2$ (the sum of squared differences
between the model and observed velocities) to find the time periods
during which the model and observed velocity profiles agree best. The
values of $\chi^2_R$ and $\chi^2_T$ are computed in $n=17$ bins in
the distance interval $R=5.5$--9.5 kpc for the radial and azimuthal
velocity profiles, respectively:

\begin{equation}
\chi^2_R= \sum^{n} \frac{(V_{R,\,\textrm{mod}}-V_{R,
\,\textrm{obs}})^2} {\varepsilon_{VR}^2},
 \label{chi2_r}
\end{equation}

\begin{equation}
\chi^2_T= \sum^{n}  \frac{(V_{T,\,\textrm{mod}}-V_{T,
\,\textrm{obs}})^2} {\varepsilon_{VT}^2},
 \label{chi2_t}
\end{equation}

\noindent  where  $\varepsilon_{VR}$ and $\varepsilon_{VR}$ are the
average uncertainties in the determination of the model and observed
velocities $V_R$ and $V_T$ in bins.  The  uncertainties in the
determination of observational   velocities $V_R$ and instantaneous
model velocities $V_R$ (obtained without averaging over the time) are
0.2 and 0.3 km s$^{-1}$, so we adopted $\varepsilon_{VR}=0.5$ km
s$^{-1}$. The uncertainties $\varepsilon_{VR}$ and $\varepsilon_{VT}$
are connected through the relation:

\begin{equation}
\varepsilon_{VT}= \frac{\kappa}{2\,\Omega} \;\, \varepsilon_{VR},
 \label{chi2_eq}
\end{equation}

\noindent where the coefficient  $\kappa/(2\Omega)$ equals 0.71 in
the interval  $R=5.5$--9.5 kpc (Fig.~\ref{dis_rat}, black line).

\begin{figure*}
\centering \resizebox{13 cm}{!}{\includegraphics{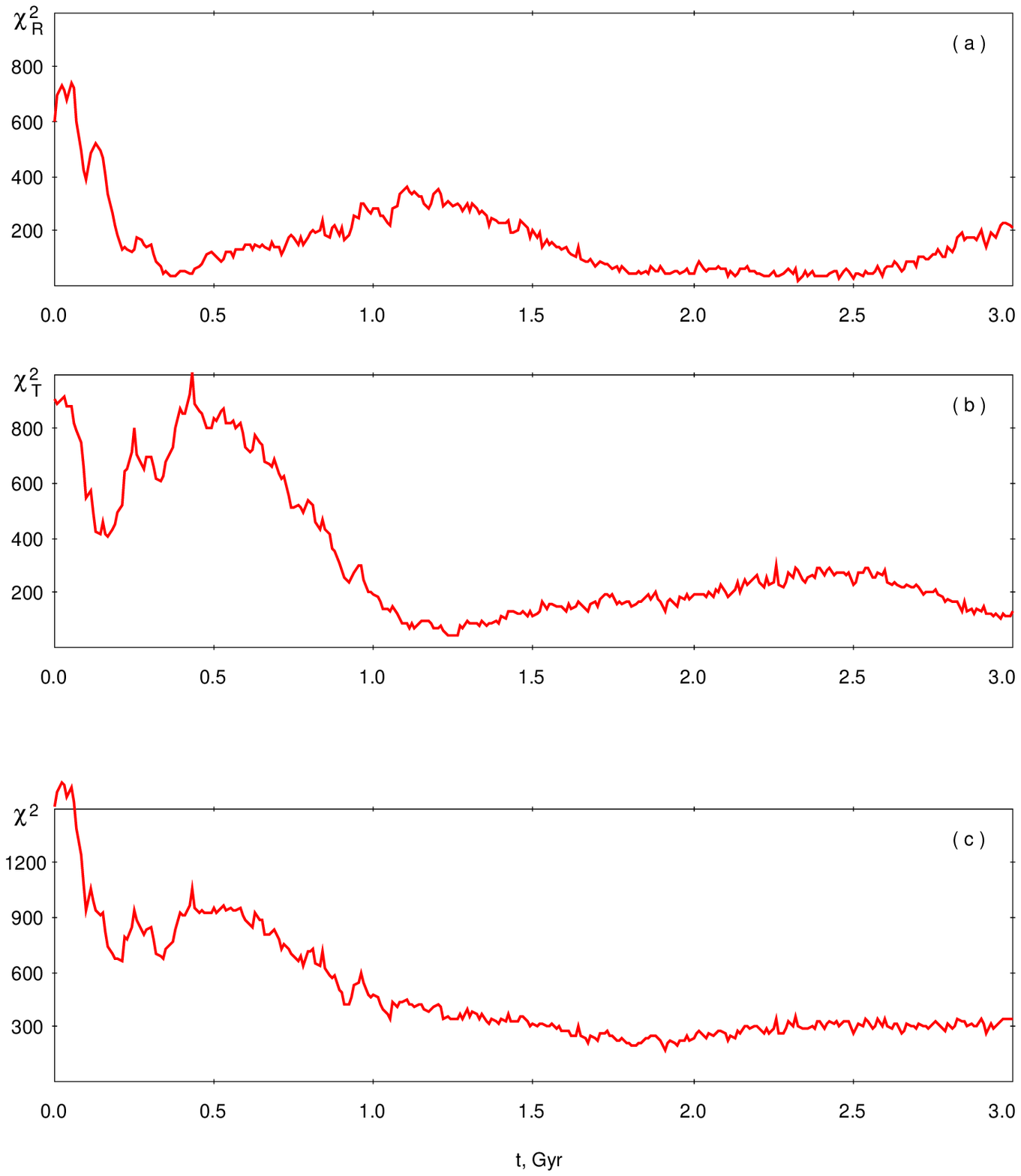}}
\caption{The functions $\chi^2_R$, $\chi^2_T$ and their sum
$\chi^2=\chi^2_R+\chi^2_T$. The $\chi^2$ function  reaches a minimum
at the time $1.8\pm0.5$  Gyr.} \label{chi2}
\end{figure*}

Fig.~\ref{chi2} shows the variations in the values of  $\chi^2_R$ and
$\chi^2_T$ as well as  in their sum $\chi^2=\chi^2_R+\chi^2_T$ with
time $t$. A step in time is $\Delta t=10$ Myr. We can see that the
$\chi^2_R$, $\chi^2_T$ and $\chi^2$ functions demonstrate small
oscillations about the average values, which are due to the
stochastic deviations of the velocities of model particles. The
$\chi^2$ function starts its decrease at $t=0.5$ Gyr and reaches a
plateau at the time 1 Gyr. The distribution of  $\chi^2$ in the time
period 1--3 Gyr is not precisely flat but exhibits a shallow minimum
at $t=1.8$ Gyr. The location of the minimum is determined with the
uncertainty of $\pm0.5$ Gyr corresponding  to $1\sigma$ probability
level.

It is just observational data derived from the {\it Gaia} EDR3
catalogue that produce the  minimum at $t=1.8\pm0.5$ Gyr. A
comparison of our model with the velocity profiles derived from the
{\it Gaia} DR2 catalogue does not yield a  minimum at the time period
1--3 Gyr.

%-----------------------    Figure  9  -----------------------------------------
%------------------------------------------------------------------------------

\subsection{Comparison of the velocity profiles calculated for different values
of $\Omega_b$ and $\theta_b$}

\begin{figure*}
\resizebox{\hsize}{!}{\includegraphics{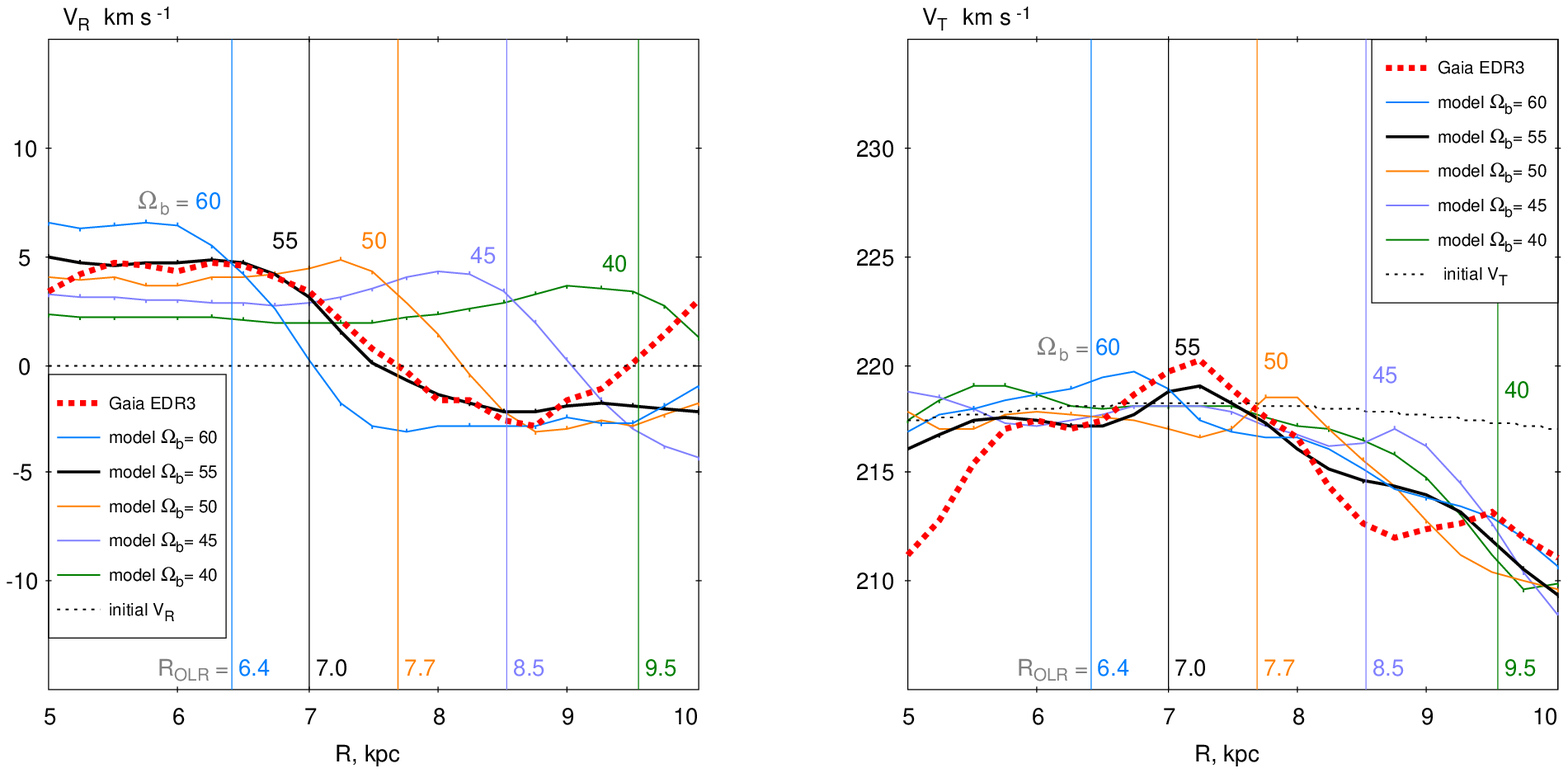}} \caption{ Model
and observed   distributions of the radial $V_R$ (left panel) and
azimuthal $V_T$ (right panel)  velocities along Galactocentric
distance $R$.  The model velocity profiles  are calculated for the
five values of the bar angular velocity: $\Omega_b=40$ (green line),
45 (violet line), 50 (yellow line), 55 (thick black line) and 60
(blue line) km s$^{-1}$ kpc$^{-1}$, and are averaged over the time
period of $t=1.5$--2.0 Gyr. The observed velocity profiles are
derived from the {\it Gaia} EDR3 data (red dashed curves). The long
vertical lines indicate the locations of the OLRs.  The upper row of
colored numbers displays  the values of $\Omega_b$ while the bottom
row gives the values of $R_{OLR}$.  The strokes on the curves
indicate the random errors in the determination of the model median
velocities in bins averaged over 50 profiles. The errors in the
determination of the observed velocities in bins are comparable with
the line width. We can see that the bar angular velocity of
$\Omega_b=55$ km s$^{-1}$ kpc$^{-1}$ provides the best agreement
between the model and observed velocity profiles while the values of
$\Omega_b=40$--50 and 60 km s$^{-1}$ kpc$^{-1}$ cause the model
velocity profiles to be shifted towards larger and smaller distances
$R$, respectively. } \label{prof_omega}
\end{figure*}

Fig.~\ref{prof_omega}  shows the distributions of the radial and
azimuthal velocities, $V_R$ and $V_T$, along Galactocentric distance
$R$ calculated for  five values of the bar angular velocity:
$\Omega_b=40$, 45, 55 and 60 km s$^{-1}$ kpc$^{-1}$. The position
angle of the bar is supposed to be $\theta_b=45^\circ$. Model
velocity-profiles are averaged for the time period  of $t=1.5$--2.0
Gyr. We can see that the value of  the bar angular velocity equal to
$\Omega_b=55$ km s$^{-1}$ kpc$^{-1}$ provides the best agreement
between the model and observed $V_R$- and $V_T$-velocity profiles.
Note, that the model velocity profiles computed, on the one hand, for
the speeds of $\Omega_b=40$, 45, 50 km s$^{-1}$ kpc$^{-1}$ and, on
the other hand, for 60 km s$^{-1}$ kpc$^{-1}$ are shifted with
respect to the observed profiles towards larger and smaller distances
$R$, respectively. The larger the bar angular velocity $\Omega_b$,
the smaller the radius of the OLR. Fig.~\ref{prof_omega} also shows
that the radial velocity $V_R$ starts decreasing nearly at the radius
of the OLR while the azimuthal velocity $V_T$ demonstrates a slight
rise just  at the distance of the OLR   and starts its drop only at
the distance exceeding the radius of the OLR by nearly 0.7 kpc: $R
\sim R_{OLR}+0.7$ kpc.

\begin{figure*}
\resizebox{\hsize}{!}{\includegraphics{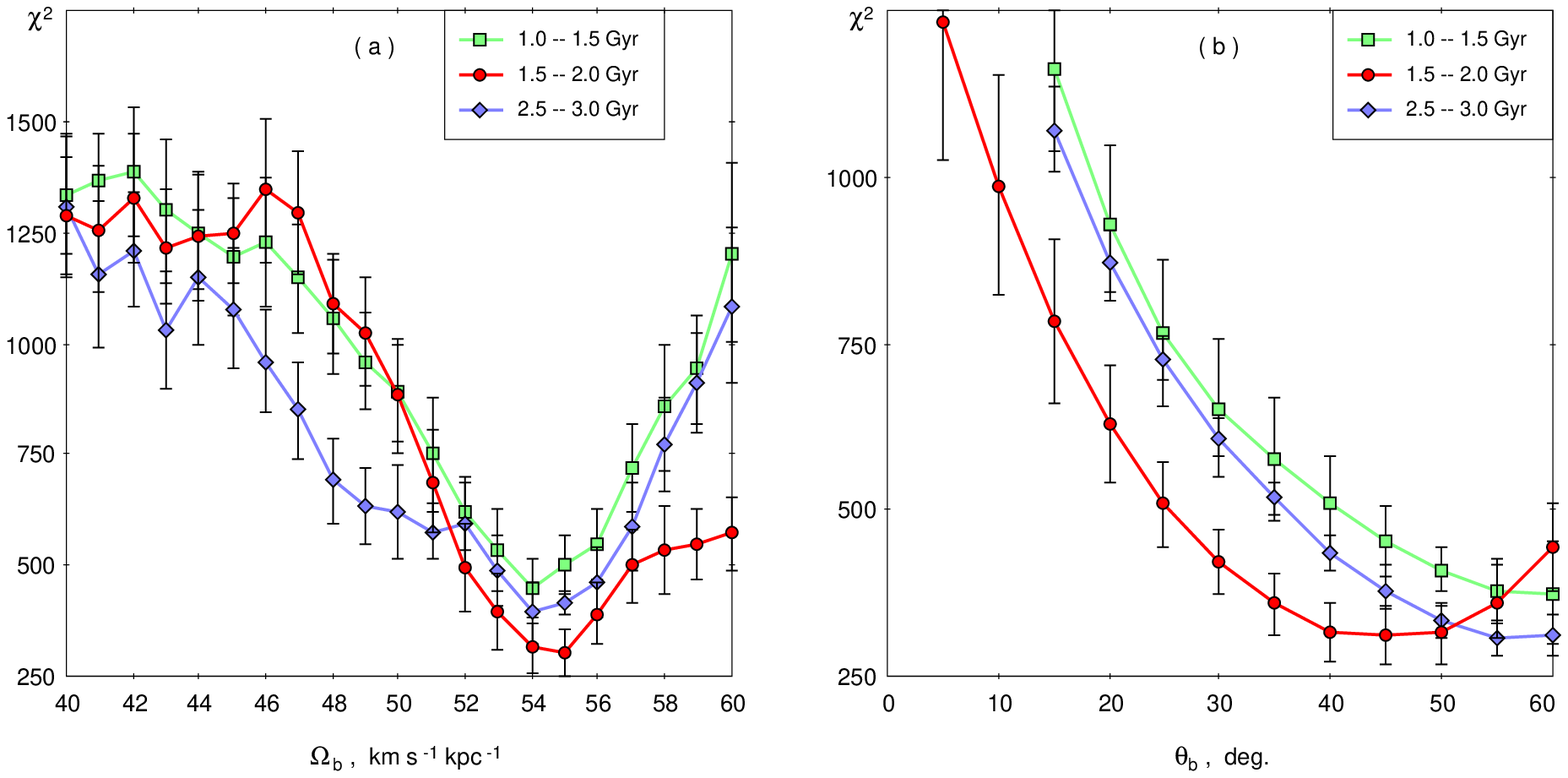}}
\caption{Variations in the   $\chi^2$ values averaged over the time
periods $t=1.0$--1.5, 1.5--2.0 and 2.5--3.0 Gyr  (a) along the
angular velocity of the bar, $\Omega_b$, and (b) along the position
angle of the bar, $\theta_b$. The vertical lines indicate the
dispersions of the $\chi^2$ values. The  $\chi^2$ function calculated
for the time period 1.5--2.0 Gyr achieves smaller values than
$\chi^2$ obtained for other time periods but the difference is close
to 1$\sigma$. (a) The minima of the $\chi^2$ functions correspond to
the value of $\Omega_b=55$ km s$^{-1}$ kpc$^{-1}$.  The 1$\sigma$
confidence interval  calculated for the time period 1.5--2.0 Gyr  is
$\Omega_b=52$--57 km s$^{-1}$ kpc$^{-1}$. (b) The $\chi^2$ function
built for the time period 1.5--2.0 Gyr demonstrates a sharp drop at
the interval 0--30$^\circ$ followed by a plateau with a shallow
minimum at $\sim 45^\circ$. The 1$\sigma$ confidence interval for the
location of the minimum is $\theta_b=25$--$60^\circ$. }
\label{omega_theta_chi}
\end{figure*}

Fig.~\ref{omega_theta_chi}(a) shows variations in  the  $\chi^2$
values averaged over the time periods  $t=1.0$--1.5, 1.5--2.0, and
2.5--3.0 Gyr as a function of the bar angular velocity $\Omega_b$.
The vertical lines indicate the dispersions of the $\chi^2$ values.
We can see that the $\chi^2$ functions achieve minima at
$\Omega_b=55$ km s$^{-1}$ kpc$^{-1}$.   The 1$\sigma$ confidence
interval  calculated for the time period 1.5--2.0 Gyr  is 52--57 km
s$^{-1}$ kpc$^{-1}$. The function $\chi^2$ computed for the time
period 1.5--2.0 Gyr achieves smaller values than $\chi^2$ obtained
for other periods but the difference is close to 1$\sigma$.

The value of the bar angular velocity of $\Omega_b=55$ km s$^{-1}$
kpc$^{-1}$ corresponds to the location of the OLR of the bar at
$R_{OLR}=7.00$ kpc. The adopted value of the solar Galactocentric
distance is $R_0=7.5$ kpc, so the radius of the OLR  must be shifted
by 0.5 kpc towards the Galactic center with respect to the solar
circle. The uncertainty in the values of $\Omega_b$, 52--57 km
s$^{-1}$ kpc$^{-1}$, produces the uncertainty in the radius of  the
OLR equal to 6.75--7.40 kpc and the shift of the OLR with respect to
the solar circle is determined with the uncertainty
$R_{OLR}=R_0-0.5_{-0.3}^{+0.4}$ kpc.

The position angle of the bar,  $\theta_b$,  is the angle between the
direction of the bar major axis and the Sun--Galactic center line.
Fig.~\ref{omega_theta_chi}(b) shows  variations in the $\chi^2$
values averaged over the time periods $t=1.0$--1.5, 1.5--2.0, and
2.5--3.0 Gyr with the position angle $\theta_b$. We can see that the
$\chi^2$ function built for the time period 1.5--2.0 Gyr demonstrates
a sharp drop at the interval 0--30$^\circ$ followed by a plateau with
a shallow minimum at $\sim 45^\circ$. The 1$\sigma$ confidence
interval for the location of the minimum is
$\theta_b=25$--$60^\circ$.

%------------------------------------------------------------------------------
%-----------------------    Figure 10  -----------------------------------------
%------------------------------------------------------------------------------
%------------------------------------------------------------------------------

%------------------------------------------------------------------------------
%-----------------------    Figure 11  -----------------------------------------
%------------------------------------------------------------------------------
%------------------------------------------------------------------------------

\subsection{Comparison between the model and observed velocity dispersions} \label{vd}

\begin{figure*}
\resizebox{\hsize}{!}{\includegraphics{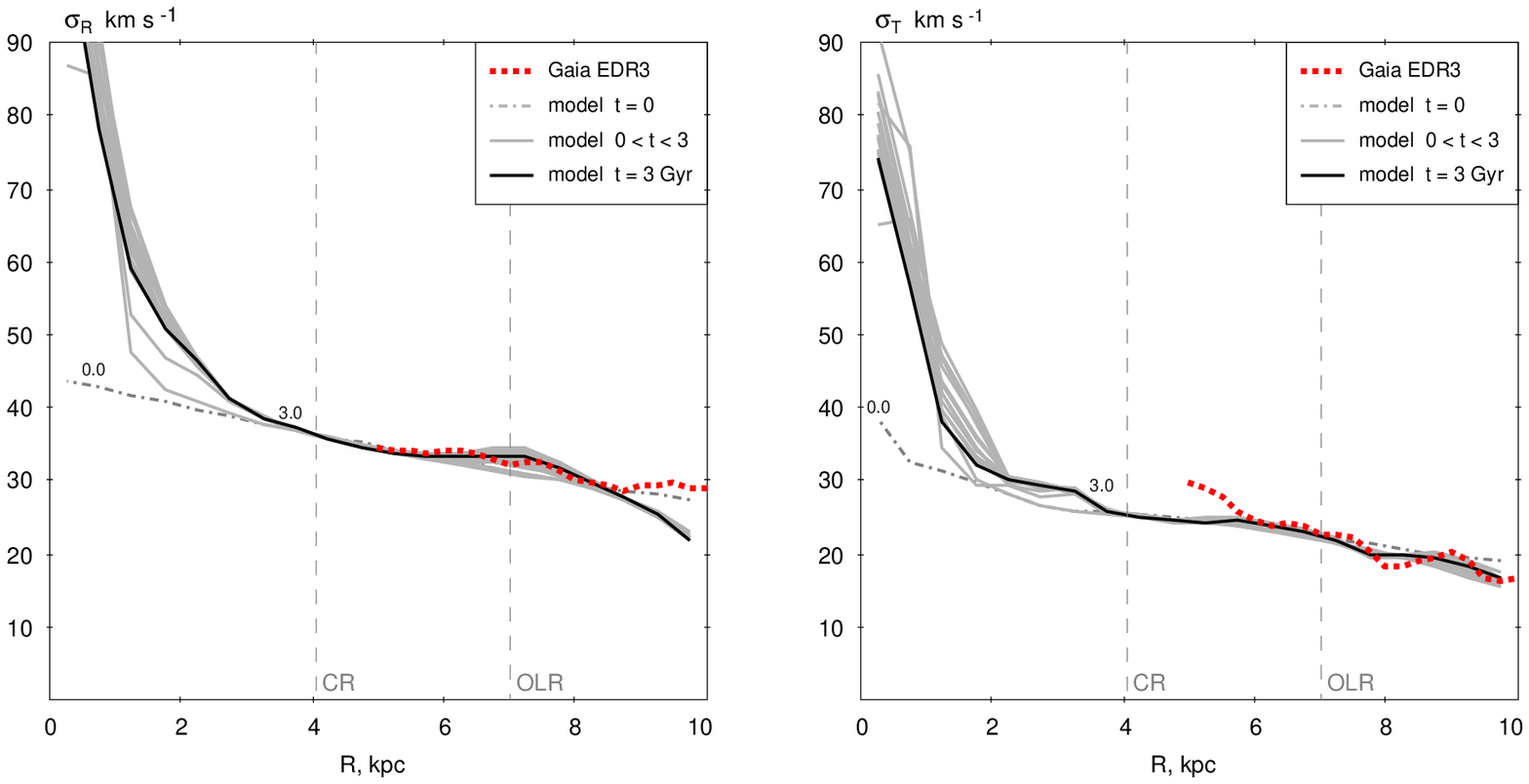}}
\caption{Variations in the radial, $\sigma_R$, and azimuthal,
$\sigma_T$, velocity dispersions as a function of Galactocentric
distance $R$ obtained for the model (black and gray curves) and
observed (red dashed curves) velocities. The model profiles of the
velocity dispersions are plotted with the time step of $\Delta t=0.2$
Gyr. The model velocity dispersions demonstrate a fast increase
inside the corotation radius $R< 4.0$ kpc during the first 0.6 Gyr
after which changes are very small. But beyond this radius,  $R>4.0$
kpc, the velocity dispersions change only slightly during the entire
simulation time. However, we can notice a small growth of the radial
velocity dispersion, $\sigma_R$, near the OLR. The model and observed
velocity dispersions agree to within $\sim 15$ per cent in the
distance interval $R=6$--9 kpc.} \label{sigma_comp}
\end{figure*}

Figure~\ref{sigma_comp} shows variations in the dispersions of the
radial, $\sigma_R$, and azimuthal, $\sigma_T$, velocities  with the
distance $R$ calculated for the model and observed velocities. The
model $\sigma_R$- and $\sigma_T$-profiles are plotted with the time
step of $\Delta t=0.2$ Gyr. We can see that the model velocity
dispersions, $\sigma_R$ and $\sigma_T$, demonstrate a fast growth
inside the corotation radius $R< 4.0$ kpc during the first 0.6 Gyr,
after which changes become very small. But beyond this radius,
$R>4.0$ kpc, the velocity dispersions change only slightly during the
entire simulation time. However, we can notice a small growth of the
radial velocity dispersion, $\sigma_R$, near the OLR: from the
initial value of $\sigma_R=31.6$ km s$^{-1}$ ($t=0$) to maximum of
34.7 km s$^{-1}$ at $t=1.3$ Gyr and then back to 32.8 km s$^{-1}$ at
$t=2.5$ Gyr. Probably, such small up-and-down changes are due to the
tuning of orbits near the resonance rings $R_1$ and $R_2$ \citep[see
discussion in][]{melnik2019}. Generally,  the model and observed
velocity dispersions agree to within $\sim 15$ per cent in the
distance interval $R=6$--9 kpc.

%-----------------------    Figure 12  -----------------------------------------
%------------------------------------------------------------------------------

\section{Discussion and conclusions}

The {\it Gaia} catalogue  mainly includes F-G-K type stars of the
main sequence of the Hertzsprung-Russell diagram
\citep{babusiaux2018} among which there are stars of different ages.
Large spectroscopic surveys such as  SDSS/APOGEE
\citep{majewski2017}, LAMOST \citep{zhao2012}  and GALAH
\citep{desilva2015} give an opportunity to obtain massive estimates
of ages of red stars in the Galaxy. The median ages of stars located
in the Galactic-midplane at the solar distance appear to lie in the
range 2--7 Gyr. In addition, the average age of disc stars increases
towards the Galactic center \citep{ness2016, frankel2019, wu2019}.
The decrease of stellar ages in the outer part of the Galactic disc
is often thought to be due to a gas accretion episode at a look-back
time of 5--9 Gyr \citep{spitoni2019, lian2020}. Another explanation
is connected with the emergence of the bar and the formation of the
outer resonance rings near the solar circle, which stimulates the
movement of gas from the outer disc to the solar neighborhood
\citep{haywood2019}.

We selected stars from  the {\it Gaia} EDR3 catalogue with  reliable
parallaxes, proper motions and line-of-sight velocities located near
the Galactic plane, $|z|<200$ pc, and in the sector of the
Galactocentric angles $|\theta|<15^\circ$. For each 250-pc wide
Galactocentrtic distance bin, we calculated the median velocities in
the radial and azimuthal directions, $V_R$ and $V_T$. The observed
distributions of the velocities over the Galactocentric distance,
$R$, have some specific features: the radial velocity, $V_R$,
demonstrates a smooth decrease from   +5 km s$^{-1}$ at the distance
of $R \approx R_0-1.5$ kpc to $-3$ km s$^{-1}$ at $R \approx R_0+1.0$
kpc while the azimuthal velocity, $V_T$, shows a sharp drop by 7 km
s$^{-1}$ in the distance interval  $R_0<R<R_0+1.0$ kpc.

The observed dispersion of radial velocities, $\sigma_R$, at the
solar distance $R_0$ is equal to $\sigma_R=31.5$ km s$^{-1}$. The
velocity dispersion $\sigma_R$ increases in the direction of the
Galactic center and its variations in the distance interval $R=5$--11
kpc can be approximated by the exponential law with a scale length of
22$\pm2$ kpc.

We built a model of the Galaxy including bulge, bar, disc and halo
components, which reproduces the observed specific features of the
velocity profiles in the Galactocentric distance interval $|R-R_0|<
1.5$ kpc. The best agreement  between the model and observed velocity
profiles corresponds to the  time $t=1.8\pm0.5$ Gyr after the start
of the simulation.

The model of the Galaxy with the bar rotating at the angular velocity
of $\Omega_b=55$ km s$^{-1}$ kpc$^{-1}$ provides the best agreement
between the model and observed velocity profiles. The  value of
$\Omega_b=55$ km s$^{-1}$ kpc$^{-1}$  sets  the OLR of the bar at the
distance of $R_{OLR}=R_0-0.5$ kpc. The 1$\sigma$ confidence interval
for the values of $\Omega_b$ is $\Omega_b=52$--57 km s$^{-1}$
kpc$^{-1}$ which corresponds to the uncertainty in the OLR location
of $\sim\pm0.4$ kpc.

The  position angle of the bar,  $\theta_b$, with respect to the Sun
corresponding to the best agreement between the model and observed
velocities is $\theta_b=45^\circ$. The 1$\sigma$ confidence interval
amounts to 25--60$^\circ$.

\section{acknowledgements}

{\small  We  thank the anonymous referee and the editor  for useful
remarks and suggestions. We also thank E.~V. Glushkova for the
discussion. This work has made use of data from the European Space
Agency (ESA) mission {\it Gaia}
(\verb"https://www.cosmos.esa.int/gaia"), processed by the {\it Gaia}
Data Processing and Analysis Consortium (DPAC,
\verb"https://www.cosmos.esa.int/web/gaia/dpac/consortium"). Funding
for the DPAC has been provided by national institutions, in
particular the institutions participating in the {\it Gaia}
Multilateral Agreement.}

\section{Data Availability}

The   data   underlying   this   article   were derived from sources
in   the   public   domain:   VizieR at
\verb"https://vizier.u-strasbg.fr/viz-bin/VizieR"

%------------------------------------------------------------------------------

\subsection{Appendix}

We  studied  the effect of observational errors onto the distribution
of the velocities $V_R$  and $V_T$  along Galactocentric distance $R$
derived from {\it Gaia} EDR3 data. Generally, the selection effects
together with errors in parallax can create some bias between the
true and observational velocities. To simulate observational errors
we created the spacial distribution of model particles close to the
observed distribution,  and added normally distributed errors to the
true values of parallaxes, proper motions and line-of sight
velocities. The standard deviations of model errors in parallaxes
($\sigma_\varpi=0.016$ mas), proper motions ($\sigma_\mu=0.015$ mas
yr$^{-1}$) and line-of-sight velocities ($\sigma_{vr}=2.0$ km
s$^{-1}$) are supposed to be equal to the average values of errors
given in the {\it Gaia} DR3 catalogue for the sample of stars
considered: $|\theta|<15^\circ$, $|z|<200$ pc, $5<R<10$ kpc.

\begin{figure*}
\resizebox{\hsize}{!}{\includegraphics{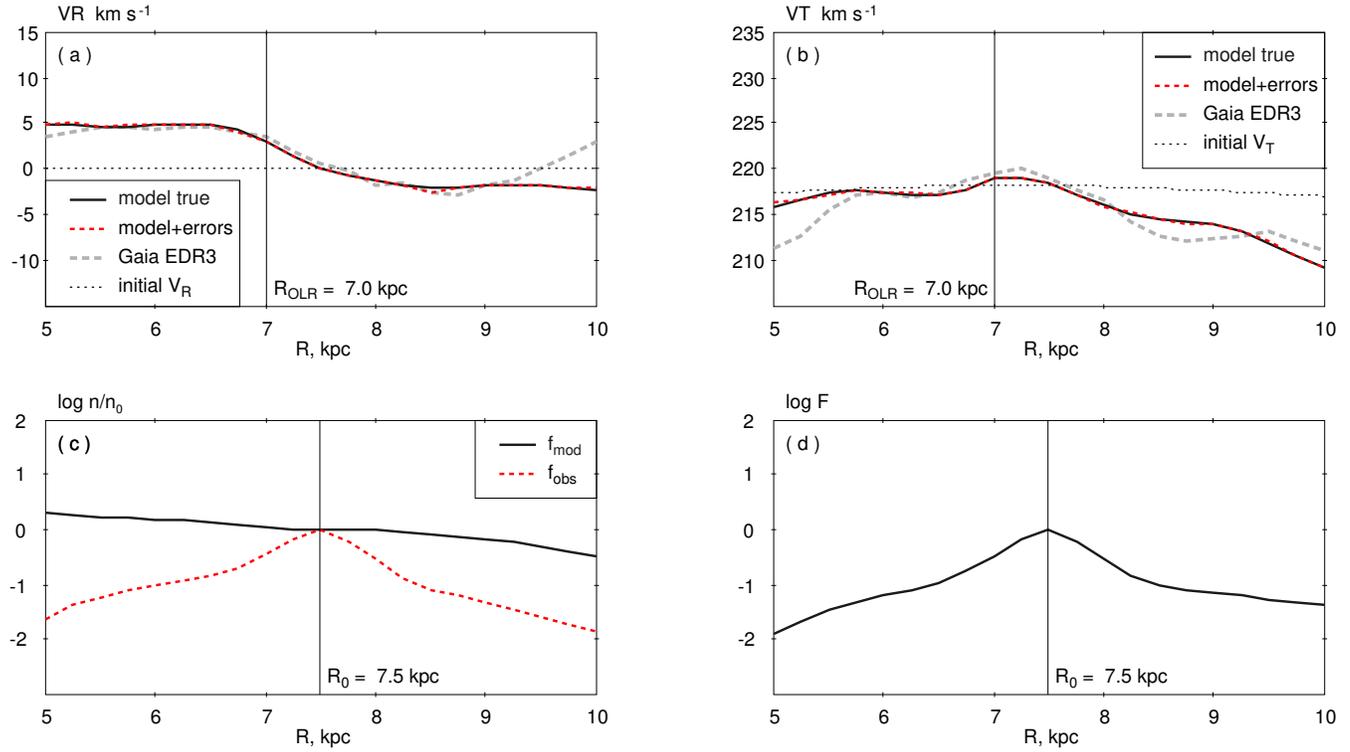}} \caption{(a, b)
Influence of observational errors  onto the  distribution of the
velocities $V_R$ (a) and $V_T$ (b) along the distance $R$. The black
solid lines describe the true  velocity-profiles of model particles
calculated in bins ($|\theta|<15^\circ$, $|z|<200$ pc, $\Delta R=250$
pc) while the red dashed lines indicate the velocity-profiles of
model particles affected by random errors in parallaxes
($\sigma_\varpi=0.016$ mas), proper motions ($\sigma_\mu=0.015$ mas
yr$^{-1}$) and line-of-sight velocities ($\sigma_{vr}=2.0$ km
s$^{-1}$). We can see that observational errors have practically no
effect on the velocity profiles in the distance interval $R=5$--10
kpc. Also shown are  the velocity profiles computed for {\it Gaia}
DR3 stars. (c, d) To study the effect of observational errors we
simulated the spacial distribution of model particles close to the
observed distribution. (c) Spacial distributions of model particles,
$n$, in bins obtained for exponential discs ($f_{mod}$) and for  {\it
Gaia} DR3 stars ($f_{obs}$) normalized at the number of particles,
$n_0$, located at the bin centered at the solar position ($R_0$=7.5
kpc). (d) Function $F=f_{mod}\;f_{obs}$ indicates the fraction of
model particles which must be retained in each bin to mimic the
selection effects in the distribution  of {\it Gaia} DR3 stars with
known line-of-sight velocities. Functions $f_{mod}$, $f_{obs}$ and
$F$ are presented in logarithmic form. } \label{mod_errors}
\end{figure*}

Figure~\ref{mod_errors} (a, b) shows the distribution of the median
velocities $V_R$  and $V_T$  calculated in  $\Delta R=250$-pc wide
bins for the true   values of $\varpi$, $\mu$ and $V_r$ and for the
values affected by observational errors. We can see that
observational errors have practically no effect on the  velocity
profiles calculated in bins at least at the distance range $R=5$--10
kpc. The average difference between the velocities calculated for the
true and observed data is 0.1 km s$^{-1}$  which is comparable to the
random errors.

Figure~\ref{mod_errors} (c, d) describes the method of modelling the
observational distribution. Figure~\ref{mod_errors} (c) shows the
spacial distributions of model particles, $n$, in bins normalized to
the number of particles, $n_0$, in the bin centered at the solar
position ($R_0=7.5$ kpc) calculated for the exponential disc
($f_{mod}$) and for {\it Gaia} DR3 stars ($f_{obs}$).  We can see
that the logarithmic distribution of model particles in the
exponential disc ($f_{mod}$) is close to the straight line while the
distribution of {\it Gaia} DR3 stars in our sample is
bell-like-shaped (see also Fig.~\ref{gaia_distrib}b). To mimic the
selection effect we  retained all model particles in the bin centered
at the solar position and only a small fraction of particles in the
bins located far from the solar circle.

Figure~\ref{mod_errors} (d) shows the fraction, $F=f_{mod}\,f_{obs}$,
of model particles which must be retained in each bin to mimic the
selection effect in the distribution  of {\it Gaia} DR3 stars with
known line-of-sight velocities \citep{sartoretti2018,  katz2019}.
Function $F$ equals unity, $F=1$, at the solar position and it is
less than 0.1, $F<0.1$, in the bins located farther than 1 kpc from
the solar circle, $|R-R_0|>1$ kpc. The asymmetry of function $F$ with
respect to the solar position means that we should exclude more
particles in the direction toward the Galactic center than in the
opposite direction.

Thus, we can neglect the observational errors in our analysis of the
velocity distributions along the distance $R$.

%-----------------------    Figure 13  -----------------------------------------
%------------------------------------------------------------------------------

%------------------------------------------------------------------------------

\end{document}